\documentclass[10pt,journal,compsoc, review]{IEEEtran}

\ifCLASSOPTIONcompsoc
  \usepackage[nocompress]{cite}
\else
  \usepackage{cite}
\fi

\ifCLASSINFOpdf
\else
\fi

\usepackage{url}
\usepackage{multirow}
\usepackage{array}
\usepackage{tikz}
\usepackage{sparklines}
\usepackage{subfig}
\usepackage{graphicx}
\usepackage{makecell}
\usepackage[utf8]{inputenc}
\usepackage{tcolorbox}

\usepackage{amsmath}
\usepackage{stfloats}

\usepackage{enumitem}
\newtcbox{\inlinecode}{on line, boxrule=0pt, boxsep=0pt, top=2pt, left=2pt, bottom=2pt, right=2pt, colback=gray!15, colframe=white, fontupper={\ttfamily \footnotesize}}

\begin{document}

\title{\Large \bf Open or Sneaky? Fast or Slow? Light or Heavy?: \\
Investigating Security Releases of Open Source Packages}

\author{Nasif~Imtiaz,
        Aniqa~Khanom,
        and~Laurie~Williams
\IEEEcompsocitemizethanks{\IEEEcompsocthanksitem Nasif Imtiaz and Laurie Williams are with the Department of Computer Science, North Carolina State University, Raleigh, NC, 27606.\protect\\
E-mail: {simtiaz, lawilli3}@ncsu.edu
\IEEEcompsocthanksitem Aniqa Khanom was with  the Department of Computer Science, North Carolina State University.}
}

\IEEEtitleabstractindextext{%
\begin{abstract}

Vulnerabilities in open source packages can be a security risk for the client projects that use these packages as dependencies. When a new vulnerability is discovered in a package, the package should quickly release a fix in a new version, referred to as \textit{security release} in this study. The security release should be well-documented and require minimal migration effort to facilitate fast adoption by the client projects. However, to what extent the open source packages follow these recommendations is not known. The goal of this study is to aid software practitioners and researchers in understanding the current practice of releasing security fixes by open source packages and identifying areas for improvement through an empirical study of security releases. Specifically, in this paper, we study (1) the time lag between fix and release; (2) how security fixes are documented in the release notes; (3) code change characteristics (size and semantic versioning) of the release; and (4) the time lag between the release and an advisory publication on Snyk or NVD (two popular vulnerability databases) for security releases over a dataset of 4,377 security advisories across seven package ecosystems (Composer, Go, Maven, npm, NuGet, pip, and RubyGems). 

We find that the median security release is available in under 4 days of the corresponding fix and contains 134 lines of code (LOC) change. Further, we find that 61.5\% of the security releases come with a release note that documents the corresponding security fix. However, Snyk and NVD may take a median of 25 days (from the release) to publish an advisory for these security releases, possibly resulting in delayed notification to the client projects. Further, security releases may also contain breaking change(s) as 13.4\% of the releases indicated backward incompatibility through semantic versioning, while 6.4\% of the releases mentioned breaking change(s) in the release notes. Based on our findings, we make four recommendations for the package maintainers and the ecosystem administrators, such as using private fork for security fixes and standardizing the practice for announcing security releases.

\end{abstract}
}
\maketitle
\section{Introduction}
Cyberattacks targeting open source packages in the software supply chain have seen a rapid surge in recent times~\cite{sonatype2021}. While modern software increasingly depends on open source packages~\cite{blackduck2021}, vulnerabilities in dependency packages can pose a security risk~\cite{cox2019surviving, zimmermann2019small}. Consequently, scrutiny from security researchers and bug bounty hunters has resulted in a continuous discovery of new vulnerabilities in the open source ecosystem~\cite{ossvuln}. 

However, detecting a vulnerability is only the start of the cure. Once a vulnerability is discovered, the ideal is that a fix will be quickly developed and be released in a new version, i.e. security release, so the client projects are able to adopt the fix. Delay in the propagation of security fixes downstream creates a window of opportunity for an attacker to exploit the unpatched vulnerability within the client projects~\cite{li2017large,chinthanet2021lags}. Therefore, packages should follow good developmental practices to ensure client projects are promptly informed about a vulnerability (e.g. through release notes or publication of a security advisory on popular vulnerability databases) and can upgrade to the security release with minimal migration effort~\cite{chinthanet2021lags, pashchenko2020qualitative, kula2018developers, martiusdoes}.

Understanding and quantifying the security release practices that are relevant to fix propagation would help (a) provide recommendations to package maintainers and ecosystem administrators;
(b) develop security metrics for dependency selection;
and (c) identify future research opportunities. An empirical measurement study is yet to be performed to understand the security release practices among open source packages, across different programming languages and ecosystems. 

The goal of this study is to aid software practitioners and researchers in understanding the current practice of releasing security fixes by open source packages and identifying areas for improvement through an empirical study of security releases. Below, we state our research questions along with the motivation behind them: 

\begin{quote}
    \textbf{RQ1: What is the time lag between a security fix and the publication of a release that includes the fix, i.e., security release?}
\end{quote}

\textbf{Motivation:} Vulnerabilities may be disclosed either publicly or privately~\cite{ruohonen2020mixed}. However, once applied to the codebase, the vulnerability fix is public by the nature of the open source. 
Even when the security aspect of a fix is not publicly discussed, research has shown that it is possible to detect security fixes through data mining approaches~\cite{ramsauer2020sound}. These fixes can give attackers information on how to design exploits. But until the fixes are included in a new release, the client projects cannot adopt the fix and remain exposed to the vulnerability, creating a ``window of opportunity'' for the attackers. Therefore, we quantitatively analyze the fix-to-release delay in the past security releases.

\begin{quote}
    \textbf{RQ2: How is the security fix documented in the release notes?}
\end{quote}

\textbf{Motivation: } Once security fixes are included in a release, vulnerability databases need to be updated and client projects need to be notified. Paschenko et al. ~\cite{pashchenko2020qualitative} find that ``well-documented, well-indicated security fixes that do not require significant development effort are more likely to be adopted by developers''. Therefore, we investigate if security releases come with a release note -- that is -- a formal document distributed with each version update that explains the notable changes in the new version~\cite{bi2020empirical, keepachangelog, changelog}. We then qualitatively analyze what information regarding the corresponding security fix is documented in the release note if one exists. 

\begin{quote}
    \textbf{RQ3: What are the code change characteristics of security releases in terms of size and semantic versioning?}
\end{quote}

\textbf{Motivation:} 
Once notified, client projects need to migrate to the fixed version of the dependency. The migration process may involve (a) reviewing new code in the update to check against supply chain attacks (e.g., injection of malicious code, backdoors)~\cite{zimmermann2019small}; (b) regression testing; and (c) changes in the client code in case of any regression. Prior research finds that developers are more likely to upgrade a package when security fixes are isolated in a separate release~\cite{pashchenko2020qualitative, martiusdoes}. However, security releases of open source packages often come bundled with unrelated functional changes~\cite{chinthanet2021lags}. Therefore, we quantitatively analyze the code changes in the security releases.  We hypothesize that the large code change indicates functional changes unrelated to the security fix have been bundled in the release. We also look at the semantic versioning (SemVer)~\cite{semver} of these releases as SemVer format is commonly used by package maintainers to indicate the type of changes that have gone into a new release and the potential migration effort it may require.

\textbf{RQ4: What is the time lag between a security release and the publication of an advisory on Snyk and NVD?}

\textbf{Motivation:} Client projects may use software composition analysis (SCA) tools, such as Dependabot~\cite{dependabot}, for notifications on dependency vulnerabilities~\cite{alfadel2021empirical, imtiaz2021comparative}. The SCA tools leverage vulnerability databases, such as NVD~\cite{nvd} or Snyk~\cite{snykdb}, to track vulnerability data among open source packages and to send notifications to the client projects. However, Imtiaz et al. ~\cite{imtiaz2021comparative} found that the outputs of different SCA tools can vary, resulting in client projects that utilize SCA tools possibly not becoming aware of a vulnerability at the time when the security release becomes available (delayed notification), or even not getting notified at all (false negative). Therefore, we quantitatively analyze the time lag between a security release and advisory publication on two popular vulnerability databases, Snyk~\cite{snykdb} and NVD~\cite{nvd}, to understand the notification delay from SCA tools that rely on these two databases.  

In this paper, we study security releases across seven package ecosystems (primary ecosystem for the corresponding programming languages): 
Composer (PHP),
Go (Go),
Maven (Java),
npm (JavaScript),
NuGet (.NET framework: C\#, F\#, and Visual Basic),
pip (Python), and
RubyGems (Ruby).
We apply a mixed-method approach to answer our research questions. 
Our \textbf{contributions} include:
\begin{enumerate}
    \item A quantitative analysis of the time lag between a security fix and the release of the fix for over 2,001 security advisories.
    \item A qualitative analysis of the release notes for 499 security releases of 350 advisories.
    \item A quantitative analysis of the code change size and SemVer versioning of security releases over 2,940 advisories.
    \item A quantitative analysis of the time lag between a security release and publication of the corresponding security advisory on Snyk and NVD, two popular vulnerability databases, over 3,655 advisories.
    \item A comparison of security release practices across seven package ecosystems.
    \item A set of recommendations for the package maintainers and the ecosystem administrators based on our findings.
\end{enumerate}

To the best of our knowledge, ours is the first study investigating RQ1, RQ2, and RQ4 for open source packages. Regarding RQ3, Chinthanet et al.~\cite{chinthanet2021lags} studied SemVer versioning and lines of code (LOC) change in security releases for the npm ecosystem, and we extend such analysis over seven ecosystems. Further, to the best of our knowledge, ours is the first study with a comparison of security release practices across packages in different languages. Code \& data for this study is publicly available~\footnote{\url{https://figshare.com/s/b0567722400c5b398c61}}.

The rest of the paper is structured as follows:
Section \ref{sec:terms} defines key terminologies used throughout this paper.
Section \ref{sec:relwork} discusses related work and establishes the motivation and novelty of our study. Section \ref{sec:dataset} explains our data collection process. Section \ref{sec:rq1}, \ref{sec:rq2}, \ref{sec:rq3}, and ~\ref{sec:advpubdelay} explain the methodology and present findings individually for the four research questions in this study. Section \ref{packageusage} discusses the impact of package usage on security release practices. In Section \ref{sec:discussion}, we discuss our findings and make recommendations for the different stakeholders of open source security. Finally, we discuss the limitations of our study in Section ~\ref{sec:threats}, before concluding our paper in Section ~\ref{sec:conclusions}.

\section{Key Terminologies}
\label{sec:terms}
In this section, we define key terminologies used throughout this paper:
\begin{itemize}
    \item \textbf{Security fix:} The code change that fixes a vulnerability. Once the fix is committed to the repository, the commit message and code change become public as per the development model of the open source.
    \item \textbf{Security release:} The release of a version that includes one or more security fixes. Client projects can then upgrade to the new release to adopt the containing security fixes.
    \item \textbf{Vulnerability database:} A database that maintains metadata on vulnerabilities, i.e. advisories. The metadata typically includes e.g., affected package, versions, vulnerability type, severity.
    \item \textbf{Security advisory:} Documentation of a vulnerability is typically referred to as security advisory~\cite{securityadvisory}. An advisory may be published before or after a security release is available. While discussing advisory publication date in this paper, we consider advisory publication on the vulnerability databases of Snyk~\cite{snykdb} and NVD~\cite{nvd}. 
    
    When a security advisory is published before a security release for the corresponding vulnerability is available, the advisory data needs to be updated once the vulnerability is fixed in a new (security) release. 
    
    \item \textbf{Release note:} A formal document distributed with each version release of a package that explains the notable changes in the new version~\cite{bi2020empirical, keepachangelog, changelog}.
    
    \item \textbf{Software Composition Analysis (SCA) tool:} SCA tools notify projects of known vulnerabilities in their open source dependencies~\cite{imtiaz2021comparative}. The tools leverage vulnerability databases to keep track of known vulnerability data.
    \item \textbf{Breaking change:} Code changes that are backward incompatible and may trigger rework for the client projects~\cite{bi2020empirical, semver}.
    \item \textbf{Semantic Versioning (SemVer):} SemVer provides a formatting guideline for version string to indicate what type of changes have gone into in the new version~\cite{semver}.
    \item \textbf{Client projects:} Client projects, i.e. downstream projects, are the projects that use an open source package as a dependency, either directly or transitively~\cite{imtiaz2021comparative}.
\end{itemize}

\begin{figure*}[]
\center
\begin{tikzpicture} [
catg/.style={draw,dashed, minimum height=2em},
thick/.style=      {line width=0.8pt}
]

\draw [thick][arrows={->[scale=2]}] (0,0) -- (15,0) ;
\draw [thick](0.8,0) -- (0.8,0.3);
\node [text width=2cm, align=center] at (0.8, 0.8) (node_a) {Vulnerability discovered};
\draw [thick](3,-0.5) -- (3,0.3);
\node [text width=3cm, align=center] at (3, 1) (node_b) {Security fix committed to the repository};


\draw [thick](7,-0.5) -- (7,0.3);
\node [text width=4cm, align=center] at (7, 1)  (node_c) {A security release is published, and a release note is distributed};
\draw [thick, dashed](13,-0.5) -- (13,0.3);
\node [text width=3cm, align=center] at (13, 1) (node_d) {A security advisory is published/ updated};

\draw [thick] [arrows={<->[scale=2]}](3, -0.3) -- (7, -0.3) node [midway, below, text width=4cm, align = center] {Fix not available to the client projects};

\draw [thick, dashed] [arrows={<->[scale=2]}](7, -0.3) -- (13, -0.3) node [midway, below, text width=5cm, align = center] {SCA tools not notifying client projects of the available security fix};

\end{tikzpicture}
\caption{A timeline on the possible windows of delay in security fix propagation.}
\label{fig:fixpropagation}
\end{figure*}
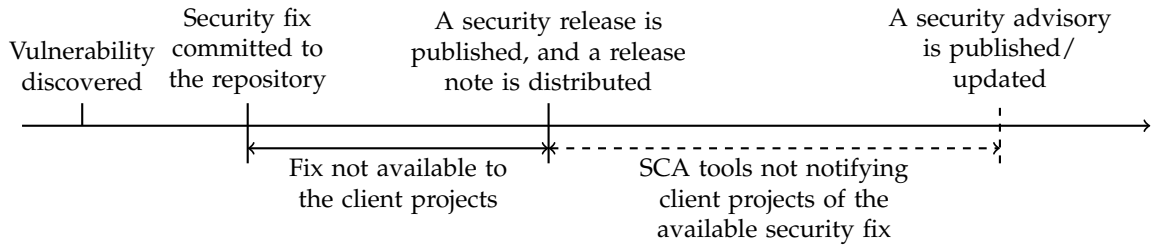

Figure ~\ref{fig:fixpropagation} shows a timeline on how security fixes in the open source packages become available to the client projects and the possible windows of delays. Note that, a client project may not upgrade to the security release even after getting notified~\cite{chinthanet2021lags, dependabot}. However, such delays are out of scope for this study, as we only investigate the security release practices from the root (of the security fix propagation) packages. Conversely, client projects may adopt a security release even when an advisory is not published. However, advisory publication delay may cause a project to remain unaware of the security fix and consequently, not adopt the new release. 

\section{Related Work}
\label{sec:relwork}
\textbf{Security risk of open source dependencies:}
Modern software projects increasingly rely on third-party open source packages as direct and transitive dependencies, as part of their supply chain~\cite{sonatype2021}. Prior work has investigated the dependency network of package ecosystems and the phenomenon of outdated dependencies
~\cite{lauinger2018thou, chinthanet2021lags, wang2020empirical, decan2017empirical, salza2019third, zerouali2019formal, gonzalez2020characterizing, stringer2020technical, decan2018evolution}. While open source dependencies in software projects bring about the benefit of code reuse~\cite{cox2019surviving}, they also come with security risks in the form of known and unknown vulnerabilities~\cite{zimmermann2019small, decan2018impact, prana2021out, hejderup2015dependencies,kula2018developers}. Therefore, managing the security risk of open source packages in the software supply chain is an area of active research~\cite{foo2019dynamics, cox2019surviving}. 

\textbf{Vulnerability Life Cycle:}
The life-cycle of a vulnerability typically refers to the time of 
i) introduction, 
ii) discovery (may never be known/ documented), 
iii) disclosure (vulnerability being disclosed to the corresponding package maintainers), 
iv) advisory publication (in vulnerability databases), 
v) fix availability, 
and vi) exploit availability~\cite{frei2006large, shahzad2012large}. 
Prior work~\cite{frei2006large,shahzad2012large, nakajima2019pilot, shahzad2019large, ruohonen2020mixed, decan2018impact, alfadel2021empirical, zerouali2021impact} has looked at the time duration between vulnerability disclosure, fix availability, and exploit availability. Li and Paxson~\cite{li2017large} have looked at the time duration between vulnerability introduction, discovery, and commit date of the code changes that fix the vulnerability.

However, the above works, while studying fix availability, did not distinguish between the code changes that fix the vulnerability, a workaround, or the release of a new version of the affected product. In the case of open source packages, client projects typically adopt the security fix by upgrading to a version where the vulnerability has been fixed~\cite{alfadel2021use, chinthanet2021lags}. However, there can be a time lag between a vulnerability fix and the subsequent release of a new version. Li and Paxson~\cite{li2017large}, while studying the commit dates of security fixes, note how the release of the fix may be delayed due to a project's release cycle, expanding the vulnerability's window of exposure. Even when the security aspect of the fixes is not publicly discussed (e.g., in the bug tracker or commit messages), prior work has identified such security fixes through data mining approaches~\cite{ramsauer2020sound}. To the best of our knowledge, ours is the first work studying the fix-to-release delay for security vulnerabilities in open source packages. 

Further, while Imtiaz et al. have studied the vulnerability reporting by SCA tools~\cite{imtiaz2021comparative}, to the best of our knowledge, ours is the first work that specifically focuses on the time lag between a security release and advisory publication, in order to understand the possible notification delay from the SCA tools (which may further expand a vulnerability's window of exposure in the downstream client projects).


\textbf{Security Fixes \& Releases:}
Prior work has investigated code changes that fix a security vulnerability~\cite{li2017large, imtiaz2019developers}. Li and Paxson ~\cite{li2017large} found security fixes are small, seven lines of code (LOC) change at the median, and tend to be localized in a single file.
Regarding releasing the fix, Chinthanet et al.~\cite{chinthanet2021lags} looked at 231 security releases across 172 npm packages and found that a) 64.5\% of the security releases are patch releases (according to SemVer format~\cite{semver}), and b) 91.8\% of the releases have other code changes unrelated to the security fix. In our work, we generalize this analysis by studying six other package ecosystems besides npm. 


\textbf{Release Notes:}
Prior work has investigated the documentation that comes with software releases. Recently, Bi et al.~\cite{bi2020empirical} have studied the release notes of 100 GitHub projects to understand what information they contain. The authors also conducted interviews and a survey to understand software practitioners' information needs from release notes. 
Their results indicate that practitioners feel the information in the release notes is scattered and vague. However, they do not look at the mention of security-related information in the release notes. Other prior works have looked at the automatic generation of release notes~\cite{moreno2016arena}, classifying issues mentioned in the release notes~\cite{abebe2016empirical}, and how system administrators consume the release notes~\cite{martiusdoes}. To the best of our knowledge, ours is the first work studying how the security fixes are documented in the release notes.

\textbf{Security Notification \& Fix Propagation:}
Once packages release a security fix, client projects will have to update to the fixed version. Kula et al.~\cite{kula2018developers} have studied why developers did not update vulnerable dependencies and found that 69\% of the developers were unaware of the security issue. While SCA tools
can be used for notifications on dependency vulnerabilities, these tools need to actively maintain their vulnerability data to ensure their reporting is accurate, complete, and up-to-date~\cite{imtiaz2021comparative}.

Bodin et al. have studied how quickly do security fixes propagate in the npm ecosystem~\cite{chinthanet2021lags}, while Paschenko et al.~\cite{pashchenko2020qualitative} have studied how security plays a role in dependency management. These studies ~\cite{chinthanet2021lags, pashchenko2020qualitative, kula2018developers, martiusdoes} point out that security releases should: i) come out as soon as the fixes have been applied; ii) be free of other functional changes; and iii) highlight the security fix to help quick adoption of the new release by the client projects. To the best of our knowledge, our is the first study measuring to what extent open source packages follow these ``best practices'' recommendations when releasing security fixes.

\section{Data set}
\label{sec:dataset}
In this section, we explain our data collection process. 

\subsection{Snyk Advisory Database \& Package Repository}
\label{sec:snykdata}

Documentation of a vulnerability is typically referred to as a security advisory~\cite{securityadvisory}.
In this paper, we work with the Snyk advisory database~\cite{snykdb}. Snyk is an SCA tool that reports vulnerabilities in open source dependencies and maintains a vulnerability database focused on open source packages~\cite{snykinfo}.
While the full Snyk database is not publicly accessible, a subset of its data remains visible on its web interface. We developed a web-scraper to obtain data from the Snyk database. Further, we complement Snyk data with data regarding individual packages fetched through APIs of the ecosystem registries.

\subsubsection{Advisory data}
\label{sec:advisorydata} Metadata on a vulnerability associated with an affected package is considered an advisory on Snyk.
Each advisory contains information on: 
(i) the affected package; 
(ii) affected versions; 
(iii) the version where the corresponding vulnerability has been fixed (referred to as security release in this study); 
(iv) vulnerability type; 
(v) description;
(vi) publication date; 
(vii) severity level (low, medium, and high); 
(viii) Common Weakness Enumeration (CWE) type(s)~\cite{cwe}; 
(ix) external reference links (including links to fix commits
(ix) a CVE identifier~\cite{cve}, if the vulnerability has been assigned an identifier. A vulnerability without a CVE identifier is referred to as a \textbf{non-CVE} in this study, as was done in prior work~\cite{imtiaz2021comparative}.

We collected data on March 5, 2021.~\footnote{Vulnerability databases are continuously updated. Our data only reflects the state of the Snyk database on March 5, 2021.} We collected 6,956 advisories across eight ecosystems: 
CocoaPods (Objective-C),
Composer (PHP),
Go (Go),
Maven (Java),
npm (JavaScript),
NuGet (.NET framework: C\#, F\#, and Visual Basic),
pip (Python), and
RubyGems (Ruby).
From this initial set,
we exclude 
i) 589 advisories (8.5\%) related to malicious packages (the malicious packages in our data set got revoked from the package registry and do not have a fixing security release, and therefore, out of scope for this study.); 
ii) advisories from the CocoaPods ecosystem (358 advisories (5.2\%), as they cover only 37 unique packages (We had 100 packages per ecosystem as a selection threshold~\footnote{After the remaining data processing steps in Section ~\ref{sec:dataset}, we had 92 packages for NuGet)}; iii) 1,233 advisories that do not have a security release listed by Snyk in our data. 

\textbf{Advisory publication date:} In this paper, we refer to advisory publication date as publication date on either Snyk~\cite{snykdb} or National Vulnerability database~\cite{nvd} (NVD).
Note that, CVEs are present in both NVD and Snyk, while non-CVEs are present only in Snyk. Therefore, for the CVEs, we take the earlier publish date among NVD and Snyk as the advisory publication date. For the non-CVEs, we take publish date on Snyk as advisory publication date.

    \begin{table*}[]
        \centering
        \caption{Study data set overview (Each RQs leverage a subset of this data set).}
        \label{tab:overview}
\begin{tabular}{lrrrrcrr}
\hline
 Ecosystem   &   Advisories &   Packages &   \makecell{Security\\ Releases} &   \makecell{\# of unique \\CWE types} & \makecell{Severity \\(low, medium, high)}           & \makecell{advisory\\with CVEs}   & \makecell{advisory with\\Non-CVEs}     \\
\hline
 Composer    &          855 &        228 &              976 &    68 & 
     \begin{sparkline}{3}
\definecolor{sparkspikecolor}{named}{gray}
\sparkspike .08 .054
\definecolor{sparkspikecolor}{named}{orange}
\sparkspike .40 .53 
\definecolor{sparkspikecolor}{named}{red}
\sparkspike .72 .43 
\end{sparkline}
 & 599 (70.1\%)          & 256 (29.9\%)  \\
  Go          &          235 &        183 &               293 &    50 & 
     \begin{sparkline}{3}
\definecolor{sparkspikecolor}{named}{gray}
\sparkspike .08 .06 
\definecolor{sparkspikecolor}{named}{orange}
\sparkspike .40 .54 
\definecolor{sparkspikecolor}{named}{red}
\sparkspike .72 .4 
\end{sparkline}  & 190 (80.9\%)          & 45 (19.1\%)   \\
 Maven       &         1,374 &        694 &              1,607 &   112 & 
     \begin{sparkline}{3}
\definecolor{sparkspikecolor}{named}{gray}
\sparkspike .08 .05 
\definecolor{sparkspikecolor}{named}{orange}
\sparkspike .40 .54 
\definecolor{sparkspikecolor}{named}{red}
\sparkspike .72 .41 
\end{sparkline} & 1,132 (82.4\%)         & 242 (17.6\%)  \\

 npm         &          792 &        540 &               858 &    82 & 
     \begin{sparkline}{3}
\definecolor{sparkspikecolor}{named}{gray}
\sparkspike .08 .05 
\definecolor{sparkspikecolor}{named}{orange}
\sparkspike .40 .47 
\definecolor{sparkspikecolor}{named}{red}
\sparkspike .72 .48 
\end{sparkline}
 & 474 (59.8\%)          & 318 (40.2\%)  \\

 NuGet       &          333 &         92 &               233 &    34 & 
     \begin{sparkline}{3}
\definecolor{sparkspikecolor}{named}{gray}
\sparkspike .08 .02 
\definecolor{sparkspikecolor}{named}{orange}
\sparkspike .40 .33 
\definecolor{sparkspikecolor}{named}{red}
\sparkspike .72 .66 
\end{sparkline}
 & 277 (83.2\%)          & 56 (16.8\%)   \\
 pip         &          567 &        269 &               534 &    84 & 
     \begin{sparkline}{3}
\definecolor{sparkspikecolor}{named}{gray}
\sparkspike .08 .06 
\definecolor{sparkspikecolor}{named}{orange}
\sparkspike .40 .62 
\definecolor{sparkspikecolor}{named}{red}
\sparkspike .72 .32 
\end{sparkline}
 & 404 (71.3\%)          & 163 (28.7\%)  \\
 RubyGems    &          221 &        121 &               311 &    47 & 
     \begin{sparkline}{3}
\definecolor{sparkspikecolor}{named}{gray}
\sparkspike .08 .03 
\definecolor{sparkspikecolor}{named}{orange}
\sparkspike .40 .59 
\definecolor{sparkspikecolor}{named}{red}
\sparkspike .72 .38 
\end{sparkline}
 & 168 (76.0\%)          & 53 (24.0\%)   \\
 \hline
 All       &         4,377 &       2,127 &              4,812 &   172 & 
     \begin{sparkline}{3}
\definecolor{sparkspikecolor}{named}{gray}
\sparkspike .08 .05 
\definecolor{sparkspikecolor}{named}{orange}
\sparkspike .40 .52 
\definecolor{sparkspikecolor}{named}{red}
\sparkspike .72 .43 
\end{sparkline}
 & 3,244 (74.1\%)         & 1,133 (25.9\%) \\
\hline
\end{tabular}
    \end{table*}

\subsubsection{Identifying package repository}
\label{sec:inferrepo}
For the analysis of the research questions in this study, we require the source code repository of the packages. We obtain metadata for each package through APIs of each of the seven ecosystem registries and collect repository URLs if present in the metadata. We used Sonatype's API~\cite{sonatypeapi} for Maven packages, while the other six package registries have their own APIs. We only considered git~\cite{git} repositories in this study~\footnote{Besides git, we only encountered subversion~\cite{svn} projects for Apache and OpenSymphony projects. We manually looked for their GitHub mirrors
and use the GitHub repository URL if we found one.}. We could not identify the repository URL for 300 packages, which further excludes 399 advisories from the collected dataset.

\subsubsection{Study dataset}
Table \ref{tab:overview} shows the final dataset used in this study. The dataset contains 4,377 advisories covering 4,812 distinct security releases from 2,217 packages. Open source packages can maintain multiple branches at the same time (e.g., 1.x.x and 2.x.x) and may push a security fix individually for all the maintained branches. Additionally, a single release can contain a fix for multiple advisories. In our dataset, 1,231 advisories (28.1\%) had more than one security release, while 738 releases (15.3\%) contained fixes for more than one advisory. The table also shows the percentage of advisories with and without a CVE identifier. 

Our advisory dataset contains 25.9\% non-CVEs which provides an opportunity to compare the characteristics between CVEs and non-CVEs. Further, over 80\% of the advisories (3,505) were disclosed on or after the year 2016 which makes our analysis representative of recent data. However, for each of the three RQs individually, we would require additional data which would further restrict our dataset, explained in the following three subsections. 

For RQ1 and RQ3, we followed a conservative automated approach in identifying fix commits for an advisory and code change in a security release, and present an accuracy analysis of the collected data. For RQ2, we performed manual analysis over a randomly sampled data subset. In the following subsections, we explain the data set for each of the three RQs.

\subsection{Data Set For RQ1}
In RQ1, we study the time lag between fix commit and the publication date of the subsequent release that includes the fix. We explain how we collected fix commits and release dates for each advisory.

\subsubsection{Identifying fix commit} \label{sec:fixcommit}
Fix commits are the commits (code changes) that fix a vulnerability. Snyk, in its external reference URLs, contains fix commit information with associated tags like `GitHub commit', `Fix Commit', `GitHub PR' pointing to the links to commit(s) and pull request(s) that have fixed the corresponding vulnerability.
To identify fix commit information for an advisory from the external reference URLs, we look for: 
(i) \textit{commit} sub-string in the tag or URL; 
and (ii) \textit{pull} or \textit{PR} sub-string in the tag and \textit{/pull/} sub-string in the URL.

For commits, the URLs contain the commit hash identifier. For pull requests, we only look at the GitHub links and use the GitHub API to list the commits under that pull request. 
We verify the validity of a fix commit if the following conditions are met:
\begin{enumerate}
    \item The commit is present in the package repository, and the reference URL points to the package repository.
    \item The commit is not present in the package repository, but the reference URL points to the package repository. Such a case arises when a commit is squashed or the corresponding branch is deleted so that the commit is no longer referenced  in the repository. However, the commit will still be viewable on GitHub~\cite{githubviewablecommit}. We verified the validity of such commits through the GitHub API.
    \item The commit is present in the package repository, but the reference URL points to a different repository. Such a case arises when the same code is shared over multiple repositories. In these cases, we considered the commit valid if the commit message in the package repository matches with the one in the reference URL, therefore, ensuring they are the same commit. 
\end{enumerate}

If the fix commit(s) is valid, we  retrieve the time of the commit. 
All time-related data in this study are converted to UTC timezone for comparison. 
Through this process, we identified fix commits for 2,497 (of 4,377) advisories. 

\subsubsection{Identifying release date} \label{sec:releasedate}
As explained in Section \ref{sec:inferrepo}, we use ecosystem registry-specific APIs to retrieve release dates, except for Maven and Go. For Maven, we crawled package version information from the Maven central repository~\footnote{\url{https://repo1.maven.org/maven2/}}. For the Go language, the packages do not publish new versions on any central registry. Instead, a new version is published by pushing a tag to the repository, and client projects can pull Go packages directly from repository URLs~\cite{gomodule}. Therefore, for Go packages, we looked at the tags in the respective package repositories and retrieved the tag annotation date where the tag matches the release version.

For the advisories we had fix commit for,
75 Maven packages that cover 225 advisories
are not hosted on Maven central repository
and therefore, we excluded them from our analysis.
From the remaining, we could not resolve the publication date for 357 security releases.
While investigating, we found the primary reason is that the fixing version is not registered in the respective package registry. While it may be technically possible to pull a package version directly from the source, we have excluded such releases in this study as client projects would not be able to  pull these releases through the package manager, which is the usual scenario. After this step, we had 2,068 advisories for analysis of RQ1.

\subsubsection{Data accuracy check}
\label{rq1validitycheck}
For 67 advisories, we found that the release date is before the fix commit date. While investigating, we found that inaccurate listing of security release version and fix commit by Snyk is the primary reason. For the rest of the 2,001 advisories, we perform a validity check over a randomly sampled set of 50 advisories by checking the correctness of (i) the fix commit, (i) the first version where the fix is included, (ii) and release date of the version. 

For 48 data points, we identified our collected data to be accurate. For 1, the fix was already included in an earlier release than the version listed by Snyk (NVD~\cite{nvd} for the corresponding CVE also has the wrong version listed). For another, the release date on Maven central repository is at a later time than the date when the corresponding version was tagged in the source repository (a possible explanation is that the package version was uploaded/updated at a later time in the Maven repository by the package maintainers). While all vulnerability databases are known to have some inaccuracies in their data~\cite{imtiaz2021comparative}, our validity check provides an indication of the accuracy of our dataset in this study. We select these 2,001 advisories to answer RQ1.        
\subsection{Data Set For RQ2}
\label{sec:dsrq2}
In RQ2, we qualitatively analyze the release notes associated with the security releases over a subset of the study data set. We randomly sample 25 CVEs and 25 non-CVEs from each ecosystem, disclosed on or after 2018. NuGet and RubyGems only have 15 and 14 non-CVEs disclosed on or after 2018 in our dataset.  Therefore, we took non-CVEs from 2017 to fill the gap. In total, we analyze 350 advisories covering 465 security releases. We focus on selecting recent security releases to capture the current trend. As the methodology for RQ2 involves manual analysis, we perform an accuracy check (and make corrections if needed) at the same time during the analysis.

\subsection{Data Set For RQ3}\label{sec:dsrq3}
In RQ3, we investigate the code change in a security release since its prior release. In this section, we explain (i) how we determine the prior release of a security release; and (ii) how we collected the source code for each release version. 

\subsubsection{Identifying prior release}
Determining the prior release of a version is different from simply chronologically ordering all the released versions, as packages can maintain multiple branches at the same time. Each package ecosystem follows a formatting algorithm to determine version ordering. APIs for Composer, npm, NuGet, and RubyGems return the version list in an ordered format according to their respective algorithms which we leveraged to determine the prior release. However, Maven and pip return the list of versions sorted in alphabetic order. We used python library, \textit{packaging}~\cite{packaging}, and \textit{mvn-compare}~\cite{mavensemver}, to sort versions for pip and Maven respectively. We found that Maven packages can contain version names not in accordance with Maven guidelines~\cite{mavenversioncompare}, e.g. \textit{1.532.2.JENKINS-22395-diag}. We discarded 234 packages that did not follow Maven naming guidelines for RQ3 analysis, as there is no reliable automatic method to determine the prior release for their security releases.~\footnote{While the security release itself may follow Maven formatting, due to the presence of at least one version that does not follow Maven formatting, we were unable to automatically determine the prior release from the full release list.}

Go packages follow SemVer~\cite{semver} formatting which we leveraged to identify prior release from the version tags available in their source repository. We excluded packages with tags that contained prefix and/or suffix strings we were unable to parse according to the SemVer formatting.  We also found and excluded packages where all metadata regarding versions prior to the security release were removed from the package registry. After this step, we could identify the prior release of 3,856 security releases over 3,520 advisories out of the 4,377 in the initial study data set.

\subsubsection{Downloading source code}
\label{dlsource} 
For each security release and its prior release version of a package, we download the source code directly from the respective package registries, except for Go and NuGet. While Go packages are not published in any central registry~\cite{gomodule}, the NuGet registry contains only the compiled artifact~\cite{nuget}. Therefore, we followed a heuristic-based approach to identify package specific code from the source code repositories of Go and NuGet packages. We explain our approach below:

We identify code for a specific version by pattern matching the version with the tag strings in the source code repositories~\footnote{Packages typically follow standard formats of putting version string as the tag name, often prefixed by 'v' (e.g. \textit{v1.0.13}). In cases, where a repository contains multiple packages that are released separately, the tag name typically also contains the package name. We only included a security release if we were able to determine the tag version following these two heuristics.} However, one repository may contain multiple packages. Go packages are named according to their exact location within the repository (e.g. \textit{github.com/go-gitea/gitea/modules/auth}), which we used to identify package specific code. NuGet packages do not contain such information in their metadata. Therefore, for NuGet packages, we apply a simple heuristic of checking if any subdirectory within a repository matches the package name. If it matches, which is often the case as per typical formatting in NuGet package repositories, we only measure the code within the subdirectory. Otherwise, we consider the full repository.


For Go and NuGet packages, we excluded security releases for which we could not identify the version tag and corresponding commit, either for the security release or its prior release versions. For the other five ecosystems, we excluded security releases if we were unable to identify a valid download URL for source code for any of the versions. After this step, overall, we could successfully download source code for 3,355 security releases (and respective prior releases) over 3,082 advisories (out of 3,520 advisories from the prior step).

\subsubsection{Measuring code change}
\label{codechangemeasure}
We measure the code change in a security release through the `git diff` tool, the default code change measuring approach in git repositories~\footnote{We ignore blank lines while measuring code change using the native git-diff feature. However, we do not ignore comment lines to keep the methodology consistent across the seven language ecosystems, as the native `git diff` tool does not have the comment line filtering feature.}. For 64 Go security releases, we found that the release does not change any file within the affected package itself. Repositories for Go packages often contain many related packages (\textit{modules,} as referred to in the Go ecosystem) that follow a single versioning scheme. Therefore, a vulnerability may get fixed in a Go module by changing code in a related module when the modules are used together. However, we exclude these 64 Go releases to avoid under-approximation in our analysis. 

We also noticed that packages often contain non-source-code files, such as documentation, configuration, auto-generated, and resource (e.g. images) files. Changes in non-source code files often inflate the code change measurement. 
Therefore, we separate the source code files and measure only source code change. We used the file formats to filter out non-source-code file changes. Our dataset includes 1,270 distinct file formats. The first author manually inspected the 182 formats for which the corresponding files were changed at least
100 times (99\% of the total file change count in our data set) and classified 42 formats as source-code files across the seven ecosystems~\footnote{We used below file formats as source code files (and exclude any other file format as non-source-code files for RQ3 analysis):
\textit{js,
ts,
java,
sh,
swift,
tsx,
h,
cc,
jsx,
php,
vue,
coffee,
c,
m,
bat,
as,
py,
patch,
ps1,
rb,
cpp,
hpp,
pl,
sql,
thrift,
cs,
go,
hx,
pm,
groovy,
scala,
asm,
jsp,
bats,
factories,
erb,
phpt,
s,
cxx,
fs,
vb,
sol.}}. Our procedure follows prior work on measuring code changes in security fixes by Li and Paxson~\cite{li2017large}.

For 136 releases, we found that the code change does not modify any source code files. Investigating 10 randomly sampled cases, we find that - (i) in 6 cases, the package made changes to its dependency versioning in the configuration files (a possible case of vulnerable dependency); (ii) in 2 cases, changes were in data files (e.g., \textit{yaml} file that stores regex patterns); and (iii) in 2 cases, the prior release already contained the fix while the security release listed by Snyk only adds a security notice in the release notes. However, we excluded configuration files as their file formats often share the same formats as auto-generated files (e.g. \textit{json}, \textit{xml}) and auto-generated files often contain large changes between releases. Further, we do not focus on fixes for vulnerable dependencies in this study.

Finally, we select 3,151 releases over 2,940 advisories for RQ3 analysis.~\footnote{We developed a python package, \textit{version-differ}, that measures the code diff between two versions of a package as per our study methodology, available at \url{https://pypi.org/project/version-differ/}.}. Figure \ref{tab:rq3exclude} shows excluded security releases and the reasons we were unable to analyze them in each of the three steps explained in this subsection.

\begin{table}[]
    \caption{Excluded security releases for RQ3 analysis}
    \label{tab:rq3exclude}
    \centering
    \begin{tabular}{rrl}
    \hline
        Advisories & Security releases &  Reason \\
        \hline
        857 & 956 & \makecell[l]{Unable to identify the prior \\release of the security release} \\
        438 & 231 & \makecell[l]{Unable to obtain  source code\\ of both the releases} \\
        142 & 204 & \makecell[l]{Source code files not modified\\ in the security release} \\
        \hline
    \end{tabular}
\end{table}

\subsubsection{Data accuracy check} We randomly sampled 10 security releases from each ecosystem to verify our collected data, which is in total 70 data points. Specifically, we verify (i) the prior release; and (ii) code change measurement by manually exploring respective source code repositories. For 67 data points, we verified our collected data to be accurate. For the other 3, file renaming and auto-generated code inflated code change measurement in 2 npm releases. In 1 npm release, the package itself is small and downloads binaries (compiled from C code)  during package installation. Therefore, while the code change in the package code is small, the effective code changes are presumably in the binaries, which we did not measure. Our validity check (96\% accuracy) provides an indication of the reliability of RQ3 findings. 

\subsection{Data Set For RQ4}
In Section \ref{sec:snykdata}, we explained how we determine advisory publication date (either on Snyk or NVD, whichever earlier). Further, in Section \ref{sec:releasedate}, we retrieved release dates for the security releases in our data set. Overall, we have security release date and advisory publication date for 3,655 advisories which we analyze to answer RQ4.  

\section{RQ1: Fix-To-Release Delay}
\label{sec:rq1}
 \begin{table*}[]
     \centering
       \caption{Analysis of time lag between a security fix and the subsequent release that includes the fix (RQ1)}
     \label{tab:timelag}
\begin{tabular}{lrrrrrrr}
\hline
 Ecosystem   &   Advisories &   Packages & \makecell{Security\\ Releases*}    & \makecell{\# of unique \\CWE types} &   \makecell{Fix-to-release \\delay\\(median days)} & \makecell{Occurrences of\\advisory published\\before fixed}   &  \makecell{Occurrences of\\advisory published\\before fix released}   \\
\hline
 Composer    &        183 &        93 & 204 (295)   &    40 &                1 & 13 (7.1\%)            & 27 (14.8\%)               \\
 Go          &        184 &       150 & 224 (242)   &    46 &                6 & 21 (11.4\%)           & 48 (26.1\%)               \\
 Maven       &        574 &       385 & 745 (807)   &    83 &               13 & 40 (7.0\%)            & 153 (26.7\%)              \\
 npm         &        553 &       412 & 586 (640)   &    72 &                0 & 72 (13.0\%)           & 113 (20.4\%)              \\
 NuGet       &         70 &        22 & 52 (75)     &    10 &                3 & 8 (11.4\%)            & 14 (20.0\%)               \\
 pip         &        324 &       200 & 308 (442)   &    68 &                6 & 39 (12.0\%)           & 79 (24.4\%)               \\
 RubyGems    &        113 &        80 & 121 (139)   &    41 &                3 & 9 (8.0\%)             & 15 (13.3\%)               \\
 \hline
 Total       &       2,001 &      1,342 & 2,240 (2,640) &   143 &                4 & 202 (10.1\%)          & 449 (22.4\%) \\
\hline
\multicolumn{8}{c}{\makecell{* A single release may contain multiple advisory fixes. 
The number in the parenthesis shows the count for \\distinct advisory and security release pairs for which we have a fix to release delay.}}
\end{tabular}
   
 \end{table*}
\begin{figure}
    \centering
    \includegraphics[scale=0.6]{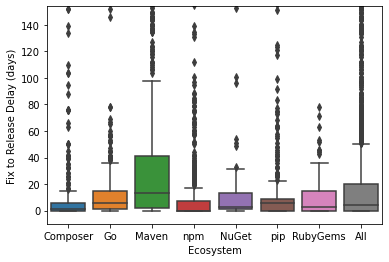}
    \caption{Box plot for fix-to-release delay for each ecosystem}
    \label{fig:timelag}
\end{figure}

In this section, we analyze the time lag between a security fix and the publication of a release that includes the fix (fix-to-release delay). For a deeper understanding, we also look at (i) how many advisories were published before a security release, and (ii) if the severity of the vulnerability impacts the fix-to-release delay, based on our collected data in Section \ref{sec:dataset}. The methodology and the findings are presented below:


\subsection{Methodology}
\label{rq1methodology}
We have fix commit information and publication date of the security release for 2,001 advisories. As one advisory may have a security release in multiple release branches, we have 2,640 distinct advisory-security release pairs for the analysis of RQ1. We identified more than one commit fixing the corresponding vulnerability for 806 advisories (38.9\%). In that case, we considered the commit date for the last of those commits as an indicator of the end of the fix. We then take the time gap between fix commit date and the publication date of the security release and present it in the unit of days. 

Besides fix-to-release delay, we also measure if an advisory was published on Snyk/NVD before a fix was made, or the fix was released, from the same motivation that an early circulation of the vulnerability gives attackers a window of opportunity.~\footnote{A vulnerability can be published as an advisory elsewhere earlier to Snyk and NVD, e.g. GitHub security advisory database. However, such a case does not change the findings on advisories that were published on Snyk/NVD before a fix or a release. Therefore, our findings here maybe only an under-approximation.} To determine the advisories published before a security release, we take the earliest release date in cases when an advisory has multiple releases in different branches. Further, we test if the severity level provided by Snyk and an advisory having a CVE identifier impacts the fix-to-release delay using Mann-Whitney U statistical test~\cite{mann1947test}. We perform Bonferroni correction~\cite{weisstein2004bonferroni} by setting $\alpha = .025$ (as we perform two hypothesis tests on the same data) to reject the null hypothesis that there lies no difference between the two tested populations.


 
 \subsection{Results}
In this section, we present our quantitative findings followed by a manual analysis over a randomly-sampled subset. 

\subsubsection{Quantitative findings}
Table \ref{tab:timelag} shows median fix-to-release time delay for all the studied open source packages and also individually across each ecosystem, while Figure \ref{fig:timelag} shows a box-plot. Across all the ecosystems, we find that the median release comes within 4 days of the corresponding security fix (0 and 20 days for the first and the third quartile, respectively). 
We find npm packages to be the quickest (50.3\% of the security releases to come under 24 hours of the fix), and Maven packages to be the slowest (only 14.9\% of the releases coming out under 24 hours). We also find Maven packages to have a larger variance of fix-to-release delay compared to the other ecosystems, as shown in Figure \ref{fig:timelag}. 

We find that advisories can be published on Snyk/NVD before a security fix is released in a new update, or can be published even before a fix has been made. As shown in Table \ref{tab:timelag}, we find that 22.4\% of the advisories were published on Snyk/NVD before the security fix was released, and 10.1\% were published even before the fix was committed. When an advisory is published before a fix is committed, we find the security release to come within 32 days of the advisory publication at a median. When an advisory is published after a fix has been made, but before the fix is released in a new version, we find the release to come within 17 days of the advisory publication at a median. The ``window of opportunity'' can be perceived to be riskier in the case where an advisory is published before the corresponding fix is released, as the advisory circulates vulnerability information. However, given that the development of open source is public by nature, a delay in releasing a security fix in any scenario carries a risk as attackers may become aware of the vulnerability by monitoring package development, but client projects may remain unaware and not take any mitigation approach, if applicable~\cite{embargo}. 

Further, 392 (19.6\%) of the advisories in our data set had fixes released in multiple branches, while 203 of them (51.8\%) of them had
all the fixing releases published on the same day. For the rest, the median time delay between the earliest and the latest releases (in different branches) was 8 days. Our data suggest that security fixes are pushed at a similar time in all the maintained branches (more than 50\% on the same day).

\subsubsection{Understanding large delays} To understand the reasons behind the time lag between a security fix and the release of the fix, we look at 50 randomly-sampled data points where the security release came at least 20 days after the corresponding fix, as such releases constitute 25\% of our dataset (the fourth quartile). We find in 6 cases that our collected data was inaccurate. In 4 cases, the fix came earlier than the version Snyk listed; in 1 case, the listed fix commit was incorrect; and in 1 case, the Maven central repository listed a date for the security release later than what appeared on the source code repository.~\footnote{While we present our accuracy check for RQ1 data in Section \ref{rq1validitycheck} (96\%), our results indicate that outlier points are more likely to have data reporting inaccuracies.} 

For the remaining 44 randomly-sampled data points, we manually explored the release date, advisory publication date, release note, and the relevant commit messages and issue-tracker discussion to understand the reasons behind the large fix-to-release delay. We present our observations below:

\begin{enumerate}
    \item \textbf{Advisory published after the security release (27):} In 27 cases, the advisory was only published once the security fix was included in a new release (10 of these advisories were published at least a year after the security release). However, for 14 of these advisories, we found that the commit message or the issue tracker discussion explicitly mentioned the security aspect of the fix. While the package maintainers may not have rushed a new release as an advisory was not yet published, such public discussion on security fixes make a vulnerability easily discoverable (e.g. by potential attackers), and validates our motivation behind RQ1.
    \item \textbf{Advisory published after a fix has been committed, but before the release (8):} In 8 cases, the advisory was published once a fix for the corresponding vulnerability was committed in the source repository, but before the fix was released in a new version. We could not identify any reasoning behind this phenomenon, besides that the package maintainers may not have prioritized the task of releasing the security fix in a new version. 
    \item \textbf{Delay in backporting the security fix (5):} In five cases, we found the security fix was included quickly only in the latest release branch of the respective package (and an advisory was published only after that). However, the delay occurred in backporting the fix to an older release branch. In this case, client projects who are not using a version from the latest branch of a package will face a delay in receiving the security fix.
    \item \textbf{Advisory published before fix (4):} In four cases, the advisory was published even before a fix was made. In two of these cases, the respective package maintainers initially suggested that it was not a vulnerability in their package, but rather a misuse of the package. In one case, the advisory was published on a security platform, presumably without giving the maintainers a time window to work on a fix. However, the maintainers responded quickly afterwards. In another case, the package maintainer mentioned that the vulnerability is only relevant in the case of legacy browsers (and therefore, not a high priority).
\end{enumerate}

Overall, we find that package maintainers may delay releasing a security fix, as they may not plan on publishing an advisory until the release anyway. However, observant attackers can still discover such security fixes through commit messages and issue-tracker discussions. 

\subsubsection{Impact of vulnerability severity}
We investigate if the severity of the vulnerability impacts the delay between a security fix and its release, with a hypothesis that the package maintainers will prioritize releasing a fix for critical vulnerabilities. To measure severity, we look at two metrics: (i) The severity level provided in Snyk advisory; and (ii) if the advisory is also published on NVD as a CVE, with an assumption that critical vulnerabilities are reported more widely (in more than one vulnerability database in this case). We present our findings below:

\textbf{Fixes for high severity advisories get released faster than medium severity ones: } We observe a statistically significant difference using the Mann-Whitney U test ($U=780357.5, p<.005$) for fix-to-release delay between high and medium severity advisories, as fixes for high severity advisories (median 3 days) were quicker than medium severity (median 5 days). We do not observe any statistically significant difference between medium and low (median 3 days) severity advisories ($U=98661.5, p>0.20$). However, only 4.5\% of the advisories in RQ1 dataset were of low severities (possibly insufficient for a comparison).

\textbf{No statistical difference in fix-to-release delay between CVEs and non-CVEs:} We found no statistically significant difference in fix-to-release delay between CVEs and non-CVEs ($U=5346.0, p >.04$). CVEs have a median fix-to-release delay of 4 days, while non-CVEs have a median delay of 3 days.

\begin{tcolorbox}
We find that open source packages include security fixes in a new release within 4 days of the fix at the median, npm being the fastest and Maven being the slowest. However, 25\% of the releases in our data set still came at least 20 days after the corresponding security fix. Such a delay creates a window of opportunity for the observant attackers, who will be informed of the vulnerability when a fix is still not available to the client projects.
\end{tcolorbox}

\section{RQ2: Documentation}
\label{sec:rq2}
We investigate if the security fixes are mentioned in the release notes for open source security releases and if they are mentioned, then what information about the security fix is present. We also investigate if other changes unrelated to the security fix are mentioned in the release note, including the inclusion of breaking changes (i.e. changes that may trigger rework for the client projects~\cite{bogart2016break, semver}). The latter two analyses (of unrelated changes and breaking changes) will help us understand the potential migration effort required by the security releases, and we can triangulate the findings from the release notes with the analysis for RQ3. Further, we also investigated if the vulnerability severity impacts if the fix would be documented in the release note or not. 

\subsection{Research Sub-Questions}
To answer this research question, we manually explore the possible locations of release notes ~\cite{keepachangelog, bi2020empirical} and identify if a security release had an associated release note. If there is a release note, we qualitatively analyze the document through open coding~\cite{khandkar2009open}. We ask the following sub-questions to answer RQ2:

\begin{enumerate}
\item How many of the security releases contain a release note? What are the sources to find the release notes?
\item How many of the release notes mention the security fix? What information regarding the security fix is present in the release notes? 
\item How many of the release notes mention changes unrelated to the security fix?
\item How many of the release notes mention breaking changes? What information regarding the breaking change is present in the release notes?
\end{enumerate}

We answer these sub-questions for 350 randomly-sampled advisories as explained in Section \ref{sec:dsrq2}. While these advisories have 465 distinct security release notes, one release can contain fixes to multiple advisories. We answer RQ2 by considering a security release for each advisory as individual data points, which gives us 499 distinct advisory-release pairs. In the following subsection, we explain our methodology to answer RQ2 for 499 data points.~\footnote{While a single release can contain multiple security advisory fixes, we present our findings for RQ2 over each advisory. Therefore, we take 499 distinct advisory-release data points as base data set for RQ2.}

\begin{table*}[]
    \centering
    \caption{Analysis on how many security releases have a release note and the release notes mention the security fix (RQ2)}
    \label{tab:manualsample}
\begin{tabular}{lrrrrrrrr}
\hline
 Ecosystem   &   Advisories\textsuperscript{1} &   Packages & \makecell{Security\\Releases\textsuperscript{2}}   &  \makecell{\# of unique \\CWE types} & \makecell{Contains\\Release \\Note}     & \makecell{Mentions\\Security\\ Fix}     & \makecell{Mentions\\Unrelated \\Changes}       & \makecell{Mentions\\Breaking\\ Change}     \\
\hline
 Composer    &         50 &        30 & 72 (98)             &    18 & 63 (64.3\%)  & 58 (59.2\%)  & 50 (51.0\%)  & 2 (2.0\%)   \\
 Go          &         50 &        49 & 65 (66)             &    25 & 54 (81.8\%)  & 43 (65.2\%)  & 43 (65.2\%)  & 10 (15.2\%) \\
 Maven       &         50 &        49 & 76 (76)             &    25 & 48 (63.2\%)  & 31 (40.8\%)  & 32 (42.1\%)  & 2 (2.6\%)   \\
 npm         &         50 &        44 & 70 (70)             &    18 & 43 (61.4\%)  & 39 (55.7\%)  & 24 (34.3\%)  & 8 (11.4\%)  \\
 NuGet       &         50 &        28 & 56 (60)             &    14 & 49 (81.7\%)  & 40 (66.7\%)  & 46 (76.7\%)  & 1 (1.7\%)   \\
 pip         &         50 &        45 & 61 (64)             &    31 & 57 (89.1\%)  & 53 (82.8\%)  & 41 (64.1\%)  & 8 (12.5\%)  \\
 RubyGems    &         50 &        40 & 65 (65)             &    20 & 49 (75.4\%)  & 43 (66.2\%)  & 24 (36.9\%)  & 1 (1.5\%)   \\
 \hline 
 All      &        350 &       285 & 465 (499)           &    68 & 363 (72.7\%) & 307 (61.5\%) & 260 (52.1\%) & 32 (6.4\%)  \\
\hline
\multicolumn{9}{c}{\makecell{1. Advisories consist of 25 CVEs and 25 non-CVEs.
2. Numbers in parenthesis indicate distinct advisory-release pair}}
\end{tabular}
\end{table*}

\subsection{Methodology}
For each security release and an advisory, we locate if there is a release note and analyze the presence of a description related to the security fix, changes unrelated to the security fix, and any breaking changes in the release note. Our methodology consists of four steps, as explained below. The first and the second author independently performed these steps for all 499 data points and resolved disagreements through discussion to finalize the manual analysis. Below we explain each of these four steps followed by an explanation of \textit{the open coding} technique and agreement rate measurement.

\subsubsection{Step I: Locating release note}
To locate if a security release contained a release note, we performed a combination of automated and manual exploratory search, divided into four steps as explained below. While the first three steps were automated and were done for all the releases, we also performed a manual search in case we failed to locate the release note automatically.
\begin{enumerate}
    \item We search for possible changelog documentation files within the source code repository. Following \textit{keep changelog}~\cite{keepachangelog} suggestions, we search for files with keywords `change', `history', `news', and `release' under documentation formats - .md, .txt, .rst, .adoc, .org, .html, .rdoc, or if the format is unspecified.
    \item The GitHub platform offers a release note option for the hosted repositories.
    We identify GitHub releases through version (and package name, if the repository contains multiple packages) string and look if the security release has a GitHub release note. 
    \item Packages can tag a commit as a release point. We look if the release tag has an associated tag message. 
    \item If we could not find a release note in the prior steps, we browse through a) external links provided in the project README file, e.g., project homepage; and/or b) package listing in the corresponding package manager site, and manually search if we can find a release note. 
\end{enumerate}

Package maintainers can put a release note in multiple locations, e.g, in both changelog and GitHub release note. In such cases, we consider the note with the longest description (e.g., GitHub release note can have an additional description on the update than what is written in the changelog). More commonly, the same note is copied in all the locations. In those cases, we give preference to the changelog, as
changelog files are stored within the source repository and linked to the other external locations. 

Moreover, the tag message or GitHub release note can be auto-generated, such as \textit{version x.y.z}, without a textual description of the changes in the release. We do not consider such cases as  a release note. If we can identify a valid note for a release, we perform the rest of the following steps.

\subsubsection{Step II: Analyzing the documentation of security fix}
Based on the vulnerability description and external reference links provided in a Snyk advisory, we classify if the release note mentions the fix for the corresponding vulnerability. To categorize what information regarding the security fix is documented, we apply open coding~\cite{khandkar2009open}. In cases where the Snyk advisory does not provide sufficient information and the release note
does not explicitly state the vulnerability,
we were unable to definitively decide if the security fix is mentioned. We classified these as unmentioned, with the rationale that a typical reader would not be able to understand that the release contains a security fix. 

\subsubsection{Step III: Identifying mention of unrelated changes}
We classify if the release note mentions any other change than the fix for the corresponding vulnerability. A release can contain multiple security fixes. However, if we found any other change than the fix for the corresponding advisory, we classified them as unrelated changes with the rationale that unless explicitly stated, we cannot independently verify if the other changes are a security fix or not. However, we keep notes when a release contains multiple security fixes, and the release note only mentions the security fixes (and no security irrelevant changes).

\subsubsection{Step IV: Analyzing the documentation of breaking change}  
We analyze if the release note mentions any breaking change that may trigger rework for the client projects, and categorize what information about the breaking change is documented through open coding.

\subsubsection{Open coding} \label{sec:opencoding} 
Open coding is a qualitative data analysis technique to create emergent concepts (codes) 
from textual data without any pre-defined concepts at the beginning of the process~\cite{khandkar2009open, hancock2001introduction}. In our study, the coders~\footnote{the first and second author of this study} independently \textit{coded} the information documented in the release notes regarding security fixes and breaking changes, which upon discussion, were grouped into categories. The coders had their first discussion after an analysis of 70 npm data points. Afterward, whenever a coder identified a possible new category, the category was discussed with the other coder.
If the category was accepted by both the coders, the data points already analyzed were revisited to identify instances of the new category. In this way, the two coders independently followed an \textit{iterative process} as suggested in prior work~\cite{wicks2017coding, pashchenko2020qualitative}. Once the two coders independently performed all the four steps mentioned above, we measure our agreement rate through Cohen's Kappa metric~\cite{viera2005understanding} for an indication of the reliability of our analysis.
Finally, we resolved the codes where we disagreed through discussion to finalize the analysis for all 499 data points.

In total, the manual analysis took approximately 50 person-hours from each of the two coders.



\begin{table*}[]
\caption{Information categories documented in the release notes regarding security fixes and breaking changes (RQ2)}
    \label{tab:codes}
    \centering
    \begin{tabular}{p{3cm}|p{5.5cm}|>{\centering\arraybackslash}p{2cm}|p{5.5cm}}
    \hline
        \makecell{\textbf{Category }\\\textbf{(count)}} & \textbf{Explanation}\textsuperscript{1} &  \makecell{\textbf{Ecosystem}\\\textbf{histogram}\textsuperscript{2}} & \textbf{Example}\textsuperscript{3} \\
        \hline
        \multicolumn{4}{c}{\textbf{How security fixes are documented}} \\
        \hline
         Security notice (198) & 
         \makecell[l]{Explicit mention of the keywords:\\ \textit{security} and \textit{vulnerability}.}
         &
         \makecell{\\
\begin{sparkline}{7}
\definecolor{sparkspikecolor}{named}{black}
\sparkspike .05 1.03 
\definecolor{sparkspikecolor}{named}{black}
\sparkspike .18 .44 
\definecolor{sparkspikecolor}{named}{black}
\sparkspike .31 .30 
\definecolor{sparkspikecolor}{named}{black}
\sparkspike .44 .4 
\definecolor{sparkspikecolor}{named}{black}
\sparkspike .57 .66 
\definecolor{sparkspikecolor}{named}{black}
\sparkspike .70 .68 
\definecolor{sparkspikecolor}{named}{black}
\sparkspike .83 .44 
\end{sparkline}}
& \textit{[SECURITY] Security release.}\\
        \hline
        Fix reference (181)&
        Link to fix commit, pull request, bug issue; commit message, pull request title; summary of fix. &
        \makecell{\\
\begin{sparkline}{7}
\definecolor{sparkspikecolor}{named}{black}
\sparkspike .05 .62 
\definecolor{sparkspikecolor}{named}{black}
\sparkspike .18 .69 
\definecolor{sparkspikecolor}{named}{black}
\sparkspike .31 .51 
\definecolor{sparkspikecolor}{named}{black}
\sparkspike .44 .73 
\definecolor{sparkspikecolor}{named}{black}
\sparkspike .57 .53 
\definecolor{sparkspikecolor}{named}{black}
\sparkspike .70 .53 
\definecolor{sparkspikecolor}{named}{black}
\sparkspike .83 .4 
\end{sparkline}}
        & \makecell[l]{\textit{Remove unnecessary escape.} \\\textit{in Regex. (\#14)}} \\
        \hline
        Vulnerability description (150) & Vulnerability type, description, severity; risk explanation; attack types.  &
        \makecell{\\
\begin{sparkline}{7}
\definecolor{sparkspikecolor}{named}{black}
\sparkspike .05 .99 
\definecolor{sparkspikecolor}{named}{black}
\sparkspike .18 .76 
\definecolor{sparkspikecolor}{named}{black}
\sparkspike .31 .36 
\definecolor{sparkspikecolor}{named}{black}
\sparkspike .44 .66 
\definecolor{sparkspikecolor}{named}{black}
\sparkspike .57 .6 
\definecolor{sparkspikecolor}{named}{black}
\sparkspike .70 .89 
\definecolor{sparkspikecolor}{named}{black}
\sparkspike .83 .7 
\end{sparkline}}
        & \textit{Fix yaml potential remote execution issue.}\\
        \hline
        \makecell[l]{Advisory \\reference (121)} & \makecell[l]{CVE identifier; link to\\ corresponding advisory.} &
        \makecell{\\
\begin{sparkline}{7}
\definecolor{sparkspikecolor}{named}{black}
\sparkspike .05 .71 
\definecolor{sparkspikecolor}{named}{black}
\sparkspike .18 .61 
\definecolor{sparkspikecolor}{named}{black}
\sparkspike .31 .61 
\definecolor{sparkspikecolor}{named}{black}
\sparkspike .44 .66 
\definecolor{sparkspikecolor}{named}{black}
\sparkspike .57 1.36
\definecolor{sparkspikecolor}{named}{black}
\sparkspike .70 1.26 
\definecolor{sparkspikecolor}{named}{black}
\sparkspike .83 .91 
\end{sparkline}}
        & \textit{Backported fix for CVE-2020-6468.} \\
        \hline
        \makecell[l]{Affected \\component (79)}&  \makecell[l]{Vulnerable API, code components\\ such as functions, classes.} &  
        \makecell{\\
\begin{sparkline}{7}
\definecolor{sparkspikecolor}{named}{black}
\sparkspike .05 .51 
\definecolor{sparkspikecolor}{named}{black}
\sparkspike .18 .46 
\definecolor{sparkspikecolor}{named}{black}
\sparkspike .31 .1 
\definecolor{sparkspikecolor}{named}{black}
\sparkspike .44 .46 
\definecolor{sparkspikecolor}{named}{black}
\sparkspike .57 .41 
\definecolor{sparkspikecolor}{named}{black}
\sparkspike .70 1.2 
\definecolor{sparkspikecolor}{named}{black}
\sparkspike .83 .87 
\end{sparkline}}
        & \makecell[l]{\textit{This fixes a security issue} \\\textit{with sequelize.json() for MySQL.}}\\
        \hline 
        \makecell[l]{Affected \\versions (33)} & \makecell[l]{List of versions affected\\ by the vulnerability.} &  
        \makecell{\\
\begin{sparkline}{7}
\definecolor{sparkspikecolor}{named}{black}
\sparkspike .05 2 
\definecolor{sparkspikecolor}{named}{black}
\sparkspike .18 .37 
\definecolor{sparkspikecolor}{named}{black}
\sparkspike .31 .61 
\definecolor{sparkspikecolor}{named}{black}
\sparkspike .44  .37
\definecolor{sparkspikecolor}{named}{black}
\sparkspike .57 .49 
\definecolor{sparkspikecolor}{named}{black}
\sparkspike .70 .12 
\definecolor{sparkspikecolor}{named}{black}
\sparkspike .83 0 
\end{sparkline}}
        & \textit{The issue applies to Rancher versions v2.0.0-v2.0.15, v2.1.0-v2.1.10, v2.2.0-v2.2.4.} \\
        \hline
        Exploit (26) & Proof-of-concept; exploit instruction; exploit condition; attack vector. & 
        \makecell{\\
\begin{sparkline}{7}
\definecolor{sparkspikecolor}{named}{black}
\sparkspike .05 2.5 
\definecolor{sparkspikecolor}{named}{black}
\sparkspike .18 .77 
\definecolor{sparkspikecolor}{named}{black}
\sparkspike .31 0 
\definecolor{sparkspikecolor}{named}{black}
\sparkspike .44 .31 
\definecolor{sparkspikecolor}{named}{black}
\sparkspike .57 .16 
\definecolor{sparkspikecolor}{named}{black}
\sparkspike .70 .16 
\definecolor{sparkspikecolor}{named}{black}
\sparkspike .83 0 
\end{sparkline}}
        & \makecell[l]{\textit{An attacker can set up a domain} \\\textit{whitelistedXexample.com that will }\\ \textit{pass the whitelist filter.}} \\ 
        \hline
        \makecell[l]{Affected \\configuration (3)} &
        Configuration required for vulnerability to manifest.
        & 
        \makecell{\\
\begin{sparkline}{7}
\definecolor{sparkspikecolor}{named}{black}
\sparkspike .05 0 
\definecolor{sparkspikecolor}{named}{black}
\sparkspike .18 1 
\definecolor{sparkspikecolor}{named}{black}
\sparkspike .31 0 
\definecolor{sparkspikecolor}{named}{black}
\sparkspike .44 0 
\definecolor{sparkspikecolor}{named}{black}
\sparkspike .57 0 
\definecolor{sparkspikecolor}{named}{black}
\sparkspike .70 2 
\definecolor{sparkspikecolor}{named}{black}
\sparkspike .83 0
\end{sparkline}}
        &  \textit{Fixed a security vulnerability for Windows users that have dcmtk installed.} \\
        \hline
        \multicolumn{4}{c}{\textbf{How breaking changes are documented}} \\
        \hline 
        Breaking change notice (22) & Explicit mention of breaking change or backward incompatibility.  &
        \makecell{\\
\begin{sparkline}{7}
\definecolor{sparkspikecolor}{named}{black}
\sparkspike .05 .27
\definecolor{sparkspikecolor}{named}{black}
\sparkspike .18 1.23 
\definecolor{sparkspikecolor}{named}{black}
\sparkspike .31 .27 
\definecolor{sparkspikecolor}{named}{black}
\sparkspike .44 .69 
\definecolor{sparkspikecolor}{named}{black}
\sparkspike .57 .17 
\definecolor{sparkspikecolor}{named}{black}
\sparkspike .70 .41 
\definecolor{sparkspikecolor}{named}{black}
\sparkspike .83 0 
\end{sparkline}}
        
        & \textit{Breaking changes: Run onSend hooks when sending a stream.}\\
        \hline
        Affected API (20) & List of breaking APIs, code components; deprecated insecure API; affected code behavior. &
         \makecell{\\
\begin{sparkline}{7}
\definecolor{sparkspikecolor}{named}{black}
\sparkspike .05 .3
\definecolor{sparkspikecolor}{named}{black}
\sparkspike .18 .9 
\definecolor{sparkspikecolor}{named}{black}
\sparkspike .31 .15 
\definecolor{sparkspikecolor}{named}{black}
\sparkspike .44 .9 
\definecolor{sparkspikecolor}{named}{black}
\sparkspike .57 0 
\definecolor{sparkspikecolor}{named}{black}
\sparkspike .70 .75 
\definecolor{sparkspikecolor}{named}{black}
\sparkspike .83 0 
\end{sparkline}}
        & \textit{--cassandra.tls is replaced by --cassandra.tls.enabled.}\\ 
        \hline 
        Action required (9) & Action required to mitigate breaking changes. &
         \makecell{\\
\begin{sparkline}{7}
\definecolor{sparkspikecolor}{named}{black}
\sparkspike .05 0
\definecolor{sparkspikecolor}{named}{black}
\sparkspike .18 .67 
\definecolor{sparkspikecolor}{named}{black}
\sparkspike .31 .33 
\definecolor{sparkspikecolor}{named}{black}
\sparkspike .44 .33 
\definecolor{sparkspikecolor}{named}{black}
\sparkspike .57 .33 
\definecolor{sparkspikecolor}{named}{black}
\sparkspike .70 1.3 
\definecolor{sparkspikecolor}{named}{black}
\sparkspike .83 0 
\end{sparkline}}
        & \textit{Those affected must upgrade to a newer Docker version.} \\
        \hline
        \makecell[l]{Code change \\reference (9)} & \makecell[l]{Link to commit, pull request\\ inducing breaking change.} & 
         \makecell{\\
\begin{sparkline}{7}
\definecolor{sparkspikecolor}{named}{black}
\sparkspike .05 .33
\definecolor{sparkspikecolor}{named}{black}
\sparkspike .18 1.67 
\definecolor{sparkspikecolor}{named}{black}
\sparkspike .31 .33 
\definecolor{sparkspikecolor}{named}{black}
\sparkspike .44 .33 
\definecolor{sparkspikecolor}{named}{black}
\sparkspike .57 0 
\definecolor{sparkspikecolor}{named}{black}
\sparkspike .70 0 
\definecolor{sparkspikecolor}{named}{black}
\sparkspike .83 0 
\end{sparkline}}
        & \textit{GPU metrics provided by kubelet are now disabled by default (\#95184).} \\
        \hline
        \makecell[l]{Affected \\configuration (7)} & \makecell[l]{Dropping support; configuration\\ change.} &
         \makecell{\\
\begin{sparkline}{7}
\definecolor{sparkspikecolor}{named}{black}
\sparkspike .05 0
\definecolor{sparkspikecolor}{named}{black}
\sparkspike .18 1.28 
\definecolor{sparkspikecolor}{named}{black}
\sparkspike .31 0 
\definecolor{sparkspikecolor}{named}{black}
\sparkspike .44 .43 
\definecolor{sparkspikecolor}{named}{black}
\sparkspike .57 0 
\definecolor{sparkspikecolor}{named}{black}
\sparkspike .70 .86 
\definecolor{sparkspikecolor}{named}{black}
\sparkspike .83 .43 
\end{sparkline}}
        & \textit{Support for the CoreOS OS distribution has been removed.} \\
        \hline
        \multicolumn{4}{l}{1. Explanation and sub-codes of the category used during open coding.} \\
\multicolumn{4}{l}{2. Histogram of relative frequency for each category, plotted in the order: Composer, Go, Maven, npm, NuGet, pip, RubyGems.}\\
\multicolumn{4}{l}{3. Excerpt from release notes exemplifying the corresponding category of information.}
    \end{tabular} 
\end{table*}

\subsection{Results}

Table \ref{tab:manualsample} shows how many security releases have a release note, and if the release note mentions the security fix, unrelated changes, and breaking changes. We identified eight categories of information
that are documented in the release notes
for security fixes and five categories of information for breaking changes. Table \ref{tab:codes} lists the categories of information on the security fixes and breaking changes along with the sub-codes used during open coding, relative frequency over the seven ecosystems, and an example for each of the categories.

Our agreement rate for these categories
ranged from 0.32 to 1.00, as shown in Table \ref{tab:kappa}. In the case of binary coding on the mention of unrelated changes, the agreement rate was 0.82. According to the interpretation of Cohen's Kappa~\cite{viera2005understanding}, the two coders had fair to perfect agreement over the different cases. Below, we discuss our findings for each of the sub-questions of RQ2.

\subsubsection{How many of the security releases contain a release note? What are the sources to find the release notes?}
We could locate a release note for 363 (72.7\%) cases in our data set. We found changelog to be the most common source (201),
followed by GitHub release note (93). In 45 cases, we located the release note on the package homepage website. In 23 cases, we considered the tag message as the release note. In 1 case, we found a separate security notice for the release. In 9 of these cases, 
we noticed the security release listed by Snyk was incorrect, as we were able to locate the fix information for the corresponding advisory in a different release.

\begin{table}[]
    \centering
    \caption{Agreement rate for categories emerged from open coding for RQ2 analysis*}
    \label{tab:kappa}
    \begin{tabular}{lr}
\hline
 Code (count)                      &    $kappa$ \\
\hline
\multicolumn{2}{c}{Categories for security fix}\\
\hline
 Security notice (198)          &    0.60  \\
 Fix reference (181)             &    0.53 \\
 Vulnerability description (150) &    0.67 \\
 Advisory reference (121)       &    0.81 \\
 Affected component (79) &    0.39 \\
 Affected version (33)  &    0.41 \\
 Exploit (26)                   &    0.32 \\
 Affected configuration (3)    &    0.5  \\
 \hline
 \multicolumn{2}{c}{Categories for breaking change}\\
\hline
 Breaking change notice (22)    &    0.83 \\
 Affected API (20)       &    0.92 \\
 Action required (9)           &    0.87 \\
 Code change reference (9)     &    1    \\
 Affected configuration (7)    &    0.72 \\
\hline
\multicolumn{2}{c}{\makecell{*In cases where the codes appeared less\\ frequently, Cohen's kappa penalizes more even for\\ small disagreements due to the unbalanced dataset}}
\end{tabular}
\end{table}

\subsubsection{How many of the release notes mention the security fix? What information regarding the security fix is present in the release notes?}
\label{sec:secfixres}

We identified the mention of security fix for the corresponding advisory in 307 cases (61.5\%). If we disregard the releases without a release note, we find that the release notes document the security fix 84.5\% of the time. We find that Python packages are most likely to mention the security fix in the release notes (82.8\%), while Maven packages are the least likely to do so (40.8\%). 

One possible reason for not mentioning security fixes in the release notes is that the omission is intentional, and the fix may be announced in a private channel. Another possible explanation is the security aspect of the fix was not recognized by the package maintainers at the time of the release.  

For the release notes where the security fix was mentioned, we identified eight categories of information to be documented about the security fix. Table \ref{tab:codes} lists the identified categories. Below, we explain the categories in the order of their frequencies. A single release note can contain multiple information categories, therefore, the total frequency of all the categories is greater than 100\%. 

\begin{enumerate}
    \item \textbf{Security notice} (39.6\%) refers to cases where the release note notifies the reader about a security fix through an explicit mention of the keyword `security' or `vulnerability'. While an explicit signal can be helpful for the client projects, we observe that release notes can be long and readers may miss the signal if not properly highlighted. 
    
    \item \textbf{Fix reference} (36.2\%) refers to cases where the release note either points to the fix commits or summarizes the fix. However, when the security is not explicitly mentioned, the readers may fail to understand that the fix addresses a potential security issue
    and not just a general bug.
    
    \item \textbf{Vulnerability description} (30.1\%) refers to cases where the release note describes the vulnerability, such as the vulnerability type and the explanation of potential risk.
    
    \item \textbf{Advisory reference} (24.2\%) refers to cases where either the CVE identifier is mentioned or an external link to a related advisory is provided. 
    We find that the majority of the advisory references are provided in the case of CVEs (89.2\%).
    
    \item \textbf{Affected component} (15.8\%) refers to the cases where the release note mentions which component of the package is affected by the vulnerability. Release notes can mention which API or function is vulnerable. Such information can be helpful for the client projects to evaluate if their projects actually use the vulnerable code of the package and prioritize fixes based on that.
    
    \item \textbf{Affected versions} (0.1\%) refers to cases where the release note states the range of versions affected by the vulnerability. Such information can be helpful for the client projects who are using a much older version of a package than the latest available and evaluate if the old version they are using is also affected by the vulnerability or not.
    
    \item \textbf{Exploit} (0.1\%) refers to cases where the release note explains how the vulnerability can be exploited. Vulnerabilities with the exploit available are deemed to be of higher risk and can help the client projects to assess how to prioritize the fixes.
    
    \item \textbf{Affected configuration} appears in only 3 release notes (0.006\%) where the release notes state the configuration in which the vulnerability is manifested. This information can also help the client projects understand whether they are affected by the vulnerability or not. 
\end{enumerate}

While all the eight categories are related to a security fix, we find 44 cases (0.9\%) where the release note only lists the fix reference and does not provide any other types of information. While we could identify the fix reference with help from the reference URLs provided with the Snyk advisory, a typical reader is unlikely to understand the security aspect of the fixes and may ignore prioritizing the update.

\subsubsection{How many of the release notes mention changes unrelated to the security fix?} 
We find that in 260 cases (52.1\%), unrelated changes to the security fix are mentioned in the release note (71.6\% of the time when we could locate a release note). Only in 91 (18.2\%) cases, the release note only mentioned the corresponding security fix and nothing else. Conversely, 48 (9.6\%) times the release note mentions unrelated changes but not the security fix. We observe that the release notes typically mention general bug fixes, new features, and enhancements besides the security fix. Further, only in 1 case, we found multiple changes to be mentioned in the release notes, and all of them were security fixes.

\subsubsection{How many of the release notes mention breaking changes? What information regarding the breaking change is present in the release notes?}

We find that in 32 (6.4\%) cases, the security releases documented breaking changes in their release notes. However, in only 5 of these cases, the breaking change was introduced due to the security fix itself. We identified five categories of information to be documented about the breaking change in the notes for security releases, as shown in Table \ref{tab:codes}. Similar to \ref{sec:secfixres}, multiple categories may appear in a single release note. Below, we explain the categories in the order of their frequencies.


\textbf{Breaking change notice} (22) refers to cases where ``breaking change'' or ``backward incompatibility'' is explicitly mentioned. In 20 cases, the \textbf{affected APIs} are listed due to the breaking change. In 9 cases, the release note mentions the \textbf{actions required} by the client projects to adjust to the breaking change. \textbf{Code change reference} (9) refers to cases where the release notes refer to the commits that have caused the breaking change. Finally, in 7 of the cases, the release notes mention the \textbf{affected configuration} under which the breaking change manifests.

\subsubsection{Impact of vulnerability severity}
We investigated if vulnerability severity impacts if and how the security fix will be mentioned in the release notes. For the comparison, we use Pearson's Chi-Squared test~\cite{plackett1983karl} with Bonferroni correction ($\alpha = .025$), and consider only the data points where we could locate a release note~\footnote{We do not measure if vulnerability severity impacts if the security release will contain a release note or not, as publishing release notes is more likely to be a package specific practice.}.

\textbf{Severity level does not impact how the security fix will be mentioned in the release notes.} We observe no statistical difference between high and medium severity advisories if the corresponding fix will be mentioned in the release notes ($X^2(1, N=343)= 0.004, p>.95$) or if the release note will provide a security notice ($X^2(1, N=343)= 2.718, p>.09$). We do not test low severity advisories, as they only constitute 27 data points in our dataset.

\textbf{CVEs are more likely to have a security notice in the release notes than non-CVEs.} For the security releases where we could locate a release note, we find no statistical difference between CVEs and non-CVEs in the likelihood of the security fix being mentioned ($X^2(1, N=363)= 3.011, p>.08$). However, we find that CVEs are more likely to come with a security notice, as categorized in this study than non-CVEs ($X^2(1, N=363)= 7.647, p<.006$). In 60.9\% of the cases, the CVEs came with a security notice in the release notes, while the non-CVEs came with a security notice only in 45.6\% of the cases. 

\begin{tcolorbox}
We find that open source packages document security fixes in a release note 61.5\% of the time. However, the security implication (of the fixes) was explicitly mentioned only in 39.6\% of the cases. The lack of proper documentation may result in client projects remaining unaware of the available security fixes. 
\end{tcolorbox}
\section{RQ3: Code Change Characteristics}
\label{sec:rq3}
We investigate the code changes in a security release and version formatting according to SemVer to estimate the potential migration effort that may be required from the client projects. Further, we also analyzed if the vulnerability severity impacts the code change size in the corresponding security release.

\begin{table*}[]
    \centering
    \caption{Analysis of code change in a security release and code change type according to semantic versioning (RQ3)}
    \label{tab:rq3}
    \begin{tabular}{lrrrrrrrrr}
\hline
 Ecosystem   &   Advisories &   Packages & \makecell{\# of unique \\CWE types}  &   \makecell{Security\\ Releases} &  \makecell{Patch \\release} & \makecell{Major \\release} & \makecell{Unstable \\release} & \makecell{\# of\\ Files\\(median)} &   \makecell{Lines of\\ Code (LOC) \\change (median)}        \\
\hline
 Composer    &           742 &          198 &    63 &          784 & 85.3\%   & 0.4\%       & 1.8\%       &              11 &         212   \\
 Go          &           145 &          109 &    42 &          164 & 62.2\%   & 1.8\%       & 23.8\%      &               3 &          59   \\
 Maven       &           589 &          293 &    77 &          754 & 66.8\%   & 1.2\%       & 3.1\%       &              13 &         326.5 \\
 npm         &           690 &          486 &    79 &          734 & 54.6\%   & 6.9\%       & 19.1\%      &               3 &          58   \\
 NuGet       &           190 &           44 &    20 &          110 & 87.3\%   & 0.9\%       & 2.7\%       &              17 &         447   \\
 pip         &           381 &          218 &    68 &          340 & 44.7\%   & 1.2\%       & 25.9\%      &               7 &         146   \\
 RubyGems    &           203 &          112 &    46 &          265 & 64.5\%   & 3.0\%       & 13.6\%      &               3 &          34   \\
 \hline
 All         &          2,940 &         1,460 &   152 &         3,151 & 66.5\%   & 2.5\%       & 10.9\%      &               6 &         134   \\
\hline
\end{tabular}
\end{table*}

\begin{figure}[]
    \centering
    \includegraphics[scale=0.6]{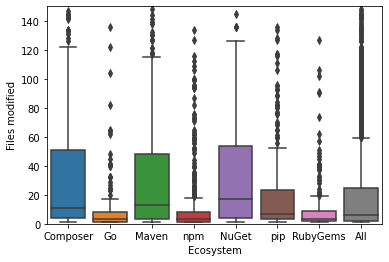}
    \caption{Box plot for the number of source code files modified in a security release}
    \label{fig:files_change}
\end{figure}

\begin{figure}[]
    \centering
    \includegraphics[scale=0.6]{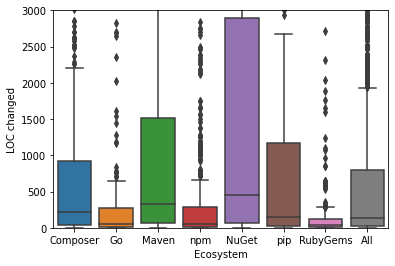}
    \caption{Box plot for the number of lines of code (LOC) changed in a security release}
    \label{fig:loc_change}
\end{figure}

\subsection{Methodology}

We measure the count of changed files and lines of code (LOC) in a security release and parse the release version to determine SemVer release type. We explain the three metrics below:

\begin{enumerate}
    \item \textbf{Files modified:} As explained in Section \ref{codechangemeasure}, we filter out the non-source-code files (e.g. configuration, documentation, resource files) and count only the source code files that have been modified in the security release.

    \item \textbf{Lines of Code (LOC) change:} For the source code files that have been modified in the security release, we take the count of lines added and removed as the total LOC change. 
    
     \item \textbf{SemVer release type:} SemVer~\cite{semver} format consists of three component numbers, X.Y.Z. All versions below 1.0.0 ($X<1$) are considered \textbf{unstable release} where developers are not expected to maintain backward compatibility for the public APIs that the package provides. Starting from 1.0.0, a change in X indicates a \textbf{major release} -- that is -- a backward incompatible release. A change in Y indicates a \textbf{minor release} where new functionalities have been added in a backward compatible manner. Finally, any change in Z indicates a \textbf{patch release} -- that is -- a release containing only backward compatible bug fixes. Therefore, releases with security fixes should typically come out as a patch release.   
     
     Further, the version can be suffixed with qualifier strings like 'pre', 'beta', 'rc' to indicate a \textbf{pre-release}. While only npm strictly enforces SemVer formatting, open source packages in the studied ecosystems typically follow this versioning format. We parsed the version strings as major, minor, patch, unstable, and pre-release according to this format. If we were unable to parse a version in SemVer format, we exclude the release from SemVer measurement.
\end{enumerate}




\subsection{Results}
In this section, we present our quantitative findings followed by a manual analysis over a randomly-sampled subset. 
\subsubsection{Quantitative findings}
Table \ref{tab:rq3} shows the median count of files and LOC change in open source security releases. The table also shows how many of the security releases were formatted as patch, major, or unstable releases as per SemVer formatting.

We find that open source security releases contain a median of 134 LOC changes, modifying a median of 6 source code files. We find RubyGems security releases to have the smallest code change, while NuGet has the highest. However, as explained in Section \ref{dlsource}, we applied a heuristic-based approach to locate package-specific source code within the repository for NuGet packages (unlike the other ecosystems where we download source code directly from the package registries), and therefore, there can be cases of over-approximation as the repository may contain non-package files as well. Figure \ref{fig:files_change} and \ref{fig:loc_change} show a box-plot for source code files modified and LOC change, respectively, in security releases across the seven ecosystems. The figures show lower variances in Go, npm, and RubyGems packages. 

In our dataset, 66.4\% of the releases were patch releases, indicating the releases only contain bug fixes. Conversely, 13.6\% of the releases indicated backward incompatibility as 2.5\% were major releases and 10.9\% were unstable releases. We find that Go, npm, pip, and RubyGems packages are more likely to have security vulnerabilities during the initial development phases (version below 1.0.0) and release fixes in the subsequent unstable releases. For the rest of the dataset, 14.5\% were formatted as minor releases indicating introduction of new functionalities while 5\% were pre-release. We observe that the pre-releases typically come before a major or a minor release. Pre-release to a major release can also indicate breaking changes are being introduced alongside a security fix, depending on when the vulnerability was introduced. 

\subsubsection{Understanding large code changes} To understand the reasons behind large code changes in security releases, we look at 10 randomly-sampled releases from each ecosystem that had more than 134 LOC changes (the median across all ecosystems). However, Go and RubyGems only had 3 releases each with LOC changes over the overall median. Therefore, we manually investigate 56 data points in total, 3 from Go and RubyGems each, and 10 each from the rest of the five ecosystems. Specifically, we look at the commits made between the two releases on a package's source repository and try to locate when the fix commit for the corresponding advisory was made through available metadata. 

In 21 cases, we were unable to identify the fix commit. However, (in these 21 cases) the changes between the two releases incorporated many commits where the commit messages indicate various types of changes and the releases appeared to follow the regular release cycle of the respective package. In 16 cases, the release came at least 7 days after the security fix commit with newer code changes and the releases appeared to follow the regular release cycle. In 16 cases, we found the release came soon after the security fix (within 7 days). However, the release included all the changes made to the package since its last release. In 1 case, the security fix itself was large in code change size. In 1 case, the release listed by Snyk (and NVD) was wrong, and the security fix was already included in an earlier release. In another case, a release was made immediately after the fix in the source repository (by tagging a new version, along with earlier changes) but was not published on the npm package registry.

Li and Paxon found security fixes to contain 7 LOC changes at median~\cite{li2017large}. However, we find the open source package releases contain 134 LOC changes at the median. Our manual analysis shows that the reason behind the large code changes in security releases is that packages bundle unrelated functional changes alongside the security fixes in these releases. Our observation is supported by prior work~\cite{chinthanet2021lags} and RQ2 findings.

Further, we observe that even if a release is published immediately after the security fix, the release is still likely to contain all the changes made to the corresponding repository branch since the last release on that branch. Therefore, the code change size of a security release may be determined by the typical release cycle and the development activity level at the time for a given package. Our findings suggest that open source packages do not follow the recommendation (from the prior work~\cite{pashchenko2020qualitative} of dedicating separate releases for security fixes.


\subsubsection{Impact of vulnerability severity}
Similar to RQ1
(the methodology explained in Section \ref{rq1methodology}),
we investigated the impact of vulnerability severity on code change size in security releases. However, a single release can contain multiple security fixes. We assign a severity level to a security release according to the highest severity vulnerability it fixes. Similarly, when measuring the difference between CVEs and non-CVEs, if a security release contains fixes for any CVE, we consider it as a CVE fix release, and otherwise non-CVE fix release. 

\textbf{No statistical difference observed for code change size of security releases between different severity levels of corresponding advisories:} Using the Mann-Whitney U test, we observe no statistical difference between security release for high (138.5 LOC, 6 files changed at the median) and medium (128 LOC, 6 files changed at the median) severity advisories and medium and low (121 LOC, 7 files changed at the median) severity advisories ($p > 0.11$ in all the cases).

\textbf{No statistical difference in code change size of security releases for CVEs and non-CVEs:} We also observe no statistical difference (using Mann-Whitney U test) in code change size between CVE and non-CVE fix releases: (i) 6 files changed at the median for both the cases, (i) 133 LOC changed in CVE releases and 138 LOC changed in non-CVE releases at the median ($p>0.19$ for both the cases).

\begin{tcolorbox}
We find that the median security release of an open source package contains 134 lines of code (LOC) changes and modifies 6 source code files. The large code changes are due to packages bundling unrelated functional changes alongside the security fix in these releases, violating the recommendation from prior work of dedicating separate releases for security fixes.
\end{tcolorbox}

\section{RQ4: Advisory publication delay}
\label{sec:advpubdelay}
In this section, we analyze the time lag between a security release and the publication of a corresponding advisory on either Snyk or NVD, two popular vulnerability databases.

\subsection{Methodology}
We have an advisory publication date and security release date for 3,655 advisories. For these advisories, we analyze RQ4 in two steps: (i) distinguish the advisories that were published before and after the security release; and (ii) quantify the advisory publication delay for the advisories that were published after the security release. 

\subsection{Results}
In this section, we present our quantitative findings followed by an analysis of RQ2 data points to understand the relationship between documentation of security fixes and advisory publication.

\subsubsection{Quantitative findings}

\begin{figure}[]
    \centering
    \includegraphics[scale=0.6]{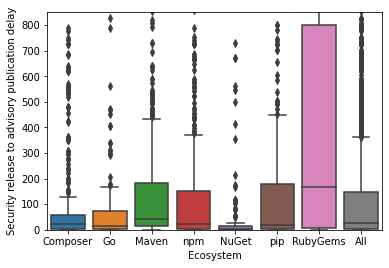}
    \caption{Box plot for the delay in advisory publication since security release}
    \label{fig:rtp_delay}
\end{figure}

Table \ref{tab:rtp_delay} shows the number of advisories that were published after a security release, and the median advisory publication delay for these advisories. Overall, 2,527 advisories (69.1\%) in our data set were published after the corresponding security release. For these advisories, we find the median delay between security release and advisory publication across all the ecosystems to be 25 days. Figure \ref{fig:rtp_delay} shows a box plot of the delays. Further, Table \ref{tab:cvenoncveadvdelay} shows a breakdown of advisory publication delay over CVEs and non-CVEs.

We find that RubyGems has a larger delay in advisory publication than the other six ecosystems. Further, 33.3\% of these RubyGems advisories are not reported on NVD as well (non-CVEs). Security researchers and SCA toolmakers may scan open source repositories to identify unreported security fixes~\cite{zhou2017automated, snykinfo}, which may be a possible reason why vulnerabilities are reported long after a fix was already released. The non-CVEs in RubyGems have a median advisory publication delay of 948 days, while CVEs have a median delay of 42 days, supporting our explanation. However, the phenomenon nonetheless shows the (higher) existence of unreported (at the time of fix) security fixes in the RubyGems ecosystem. Overall, 73.2\% of the 2,527 advisories (that were published after the security release) are CVEs. While CVEs have a median advisory publication delay of 21 days, non-CVEs have a median delay of 55 days. We found the difference to be statistically significant through the Mann-Whitney U test ($U=464128.0, p <.0001$). 

Similarly, we find that high severity advisories have a significantly lesser advisory publication delay (20 days at the median) than the medium severity (33 days at the median) advisories ($U=617888.5, p < .0001$). We only have 108 low severity advisories in our RQ4 dataset, with a median delay of 25 days, and no statistically significant difference with the high or medium severity advisories.   

While advisories published before a security release cause a vulnerability to have widespread circulation without having a fix available, the long delay in advisory publication after the security release suggest that: (i) there may be unknown security risks in using outdated dependencies~\cite{decan2018evolution}, as some security fixes may not have been reported (yet) to a vulnerability database; and (ii) open source packages need to employ standardized practices when announcing security releases, such as documenting in the release notes.

\begin{table}[]
    \centering
    \caption{Analysis of time lag between security release and advisory publication (either on NVD or Snyk)}
    \label{tab:rtp_delay}
\begin{tabular}{lrrr}
\hline
 Ecosystem  & Advisories & \makecell[c]{Advisories\\ published\\ after security\\ release}           &   \makecell[c]{Median\\ Advisory\\publication \\delay*\\(days)} \\
\hline
 Composer    &  751 & 509 (67.8\%)  &          21 \\
 Go          &  211 & 145 (68.7\%)  &          15 \\
 Maven       &  937 & 693 (74.0\%)  &          43 \\
 npm         &  735 & 540 (73.5\%)  &          24 \\
 NuGet       &  241 & 150 (62.2\%)  &           2 \\
 pip         &  569 & 367 (64.5\%)  &          19 \\
 RubyGems    &  211 & 123 (58.3\%)  &         167 \\
 \hline
 All         & 3,655 & 2,527 (69.1\%) &          25 \\
\hline
\multicolumn{3}{c}{* For the advisories published after security release}
\end{tabular}
\end{table}

\begin{table}[]
\caption{Comparison of advisory publication delay (median days) between CVEs and Non-CVEs that were published after the security release}
    \label{tab:cvenoncveadvdelay}
    \centering
    \begin{tabular}{lrrrr}
    \hline
        Ecosystem & CVEs & \makecell[c]{Publication\\ delay for\\ CVEs} & Non-CVEs & \makecell[c]{Publication\\ delay for\\ Non-CVEs} \\
        \hline
 Composer    &             355 &            18 &             154 &          21   \\
 Go          &             121 &            13 &              24 &          33.5 \\
 Maven       &             601 &            41 &              92 &         100   \\
 NuGet       &             134 &             2 &              16 &         127   \\
 RubyGems    &              82 &            42 &              41 &         948   \\
 npm         &             307 &            13 &             233 &          56   \\
 pip         &             250 &             7 &             117 &         187   \\
\hline
        All & 1,850 & 21 & 677 & 55 \\
        \hline
    \end{tabular}
    
\end{table}

\subsubsection{Security fix mentioned in the release note but delay in advisory publication}

In RQ2, we qualitatively analyzed release notes of the security releases. From that data, we investigate the relation between the delay in advisory publication and if/how security fixes were documented in the release notes. We have 343 distinct advisory-security release pairs for whom (i) we manually analyzed the release notes (in RQ2); (ii) we have the publication date of the release; and (iii) the advisory was published after the security release. We make the following observations based on these 343 data points:

\textbf{The security fixes that were not mentioned in the release notes have a median advisory publication delay of 161 days:} The 86 security releases where we could not locate a release note have a median advisory publication delay of 21.5 days. However, the 44 releases where we could locate a release note, but the note does not mention the security fix, have a median advisory publication delay of 161 days. A possible explanation is that the security aspect of the fixes was not known at the time, and the vulnerability was only recognized months later.

\textbf{Notification to the client projects may get delayed even when the security fix is explicitly mentioned in the release notes:} The 213 security releases where the release note mentions the security fix have a median advisory publication delay of 13 days. However, in 25\% of these cases, the advisory was published at least 54 days after the security release, potentially creating a lag between the time of a security release and the time the client projects are notified. Further, 146 of these releases that explicitly mentioned security or vulnerability in the release notes (\textit{security notice} category from RQ2) have a median advisory delay of 9 days. However, in 39 of these cases (26.7\%), the advisory was published at least 30 days after the security release.

\begin{tcolorbox}
We find that there can be a time lag between a security release and an advisory publication, resulting in client projects receiving delayed notification of the corresponding vulnerabilities. Such a delay further expands the window of opportunity for the attackers discussed in RQ1, as client projects may still remain unaware of the known vulnerabilities and the available security releases.
\end{tcolorbox}

\section{Impact of package usage on security release practices}
\label{packageusage}
We measured how vulnerability severity impacts security release practices when reporting findings for our research questions. Besides vulnerability severity, another hypothesis can be that the package with a higher usage (count of client projects, i.e. dependents) will employ better practices when releasing security fixes. To investigate the impact of package usage, we need to know the number of dependents of a package at the time of the security release. However, such historical data is not available for any of the seven ecosystems (to the best of our knowledge). Therefore, we were unable to measure package usage impact when discussing the findings of our research questions.

However, we can look at the current number of dependent projects of a package (as an estimator of the package usage at the time of security release), and measure if packages with higher dependents have a lower fix-to-release delay, lower code change size, better documentation practices, and lower advisory publication delay when releasing security fixes. We obtain the count of dependent project repositories for each package across all the ecosystems through statistics available on \textit{libraries.io}~\cite{jeremy_katz_2020_3626071}. We obtain the package statistics during September 2021. However, the statistics obtained from \textit{libraries.io} is likely to be an under-approximation, as the dependent count does not include (i) closed-source repositories; and (ii) repositories not tracked by \textit{libraries.io} (e.g. repositories outside GitHub). Further, the measurement may not be consistent across different ecosystems as well.

Table \ref{tab:packageusage} shows the Spearman's rank correlation~\cite{sedgwick2014spearman} for fix-to-release delay, LOC change, and advisory publication delay for security releases with the count of dependent projects.
Overall, we find that there exists a statistically significant negative correlation between the count of dependent projects in case of fix-to-release delay ($r=-0.10$) and LOC change ($r=-0.09$) in security releases, an inverse relation as hypothesized. However, the correlation coefficient is weak according to the typical interpretation~\cite{akoglu2018user}. We do not find a statistically significant correlation in the case of advisory publication delay. 

Similarly, we compared if packages that mentioned a security fix in the release notes and the ones that did not, differ in the number of dependent packages and repositories through the Mann-Whitney U test~\cite{mann1947test}. However, we did not find a statistical difference for dependent packages or for dependent repositories between the two populations ($p > 0.40$ in both cases). Overall, our analysis suggests that using a popular package does not ensure that the package will follow good security releases practices. 

\begin{table}[]
    \centering
    \caption{Impact of package usage on fix-to-release delay, LOC change, and advisory publication delay of security releases}
    \label{tab:packageusage}
    \begin{tabular}{lrrr}
         Ecosystem & \multicolumn{3}{c}{\makecell{Correlation of dependent\\ projects with security releases'}}\\
         \hline
          & \makecell{Fix-to-release\\ delay}& \makecell{LOC\\change} & \makecell{Advisory \\publication\\ delay}\\
          \hline
          Composer & *0.10 & *-0.15 & *0.24\\
          Go & -0.07 & 0.07 & 0.19\\
          Maven & *-0.11 & 0.07 & 0.02 \\
          npm & *-0.08 & 0.00 & *0.08\\
          NuGet & *0.05 & 0.00 & 0.03 \\
          pip & *-0.30 & 0.04 & 0.14\\
          RubyGems & *-0.15 & *-0.18 & -0.11\\
          \hline
          All & *-0.10 & *-0.09 & -0.02 \\
         \hline
         \multicolumn{4}{c}{* indicates statistically significant correlation with $p < 0.01$}
    \end{tabular}
\end{table}

\section{Discussion}
\label{sec:discussion}

In this section, we discuss (i) the possible cases of delay in downstream propagation of security fixes; (ii) the possible regression risk for the client projects in adopting the security releases; (ii) a comparison of security release practices between CVEs and non-CVEs; and (iii) a comparison of security release practices between the seven studied ecosystems. Afterward, based on our findings in this paper, we make four recommendations for the package maintainers and the ecosystem administrators. 

\subsection{Delay in fix propagation}

We have found the time lag between a security fix and the release that includes the fix to be four days at the median, which may be deemed as reasonable given the developmental activities required before a release, such as maintainer coordination, testing, etc. We discuss how security fixes may be kept private until release in Section \ref{sec:recommendation} to reduce the risk of fix-to-release delay. 
However, a security release itself does not ensure the security of the client projects. Client projects need to be informed of the available security release, and upgrade to the new version to adopt the security fix.
We find that the release notes of the security releases lack documentation 38.5\% of the time. Further, there can be a large delay in advisory publication as well. Such delays expand a vulnerability's exposure and increase the risk of security breaches, such as the Equifax data breach~\cite{equifax}. 

While SCA tools like Dependabot are the primary means for the client projects for notification on upstream security fixes, prior work has found that SCA tools suffer from both false positives and false negatives~\cite{imtiaz2021comparative}. In this paper, we have found that there can be a time lag between a security release and the corresponding advisory publication, causing SCA tools like Snyk to send delayed notifications. Furthermore, we have observed that the security advisories may not contain information on the available security release as well. In our collected data from Snyk, 1,413 security advisories (20.3\% of 6,956 advisories) did not have a security release listed. When manually examining 25 randomly-sampled advisories without a security release, we found that in 13 of these cases, a security release is actually available, but Snyk lists all versions of the affected package as vulnerable~\footnote{For the rest, in 7 cases, the affected packages are unmaintained. In 3 cases, a fix was made but was not released in a new version yet. In 2 cases, we were unable to identify if a fixing version exists or not.}. Inaccurate data, therefore, can result in SCA tools sending false vulnerability alerts to the client projects~\cite{imtiaz2021comparative}.

Overall, reporting vulnerability data to advisory databases and subsequent updates of the data with information such as security releases primarily involves a manual reporting process. Due to this non-automated process, security releases may remain unreported. As a result, client projects, who are not using the latest versions of their dependencies, may remain unaware of vulnerabilities even when using an SCA tool. Conversely, clients can keep receiving false alerts even when they are using a fixed version. In Section \ref{sec:recommendation}, we discuss how employing standardized practices while announcing security releases can help us automate the notification process to the client projects at the time of the security release.

\subsection{Regression risk when upgrading to a security release}

Dependency updates may introduce changes that cause the client projects' code to stop working or to have deteriorated performance. Prior research has found that such regression risk is a major reason developers do not want to update their dependencies, even when they are aware of existing vulnerabilities~\cite{pashchenko2020qualitative, kula2018developers}. The analysis in RQ2 and RQ3 are partially motivated with an objective to estimate the regression risk carried by security releases. Based on our findings, we present our observations:

\textbf{Security-critical applications should refrain from using unstable package releases as dependencies:} Packages with version below 1.0.0 are considered not stable (according to SemVer formatting), and the package maintainers are not expected to maintain backward compatibility when making new changes. Therefore, when a vulnerability is discovered in an unstable package, the security release may also come with some breaking changes. In our dataset, 10.9\% of the security releases were unstable releases, primarily coming from Go, npm, pip, and RubyGems packages. In the python ecosystem (pip), 25.9\% of the security fixes came in packages that are still at an unstable release. Of the 32 security releases where we identified the mention of breaking change in the release notes~\footnote{All 32 cases where we identified mention of breaking change in the release notes, are distinct releases.}, 5 are unstable releases. Packages may go through rapid development before a stable version is declared, and there may be a higher likelihood of (introduction of) vulnerabilities at this stage. Furthermore, the security fix may come bundled with backward incompatible changes, which necessitates changes in the client code as well. Therefore, security-critical applications should avoid the possible security hazards of using unstable releases of open source packages.

\textbf{Packages under active development do not (typically) dedicate a separate release for security fixes:}
Based on developer interviews, Paschenko et al.~\cite{pashchenko2020qualitative} recommend packages to
dedicate separate releases for security fixes (i.e. not including unrelated functional changes in the same release) to help fast adoption by the client projects, as developers may view functional changes to be a potential avenue for introducing a regression. Chinthanet et al.~\cite{chinthanet2021lags} have analyzed 231 security releases from the npm ecosystem and found that the median release contains 219 lines of code (LOC) change~\footnote{In our study, we found that the median npm security release contains only 58 LOC change. ~\cite{chinthanet2021lags} measured code changes through comparing releases on GitHub repositories which may cause an over-approximation in the measurement as repositories often contain multiple packages or test files that are not included in the published package.}. However, upon manually analyzing the commits, they identified that only 10 LOC changes (at a median) in the releases were related to the security fix.

In our work, we found that the median open source security release contains 134 LOC changes. In our manual analysis (in RQ2 and RQ3), we found that the releases may contain code changes unrelated to the security fix. We observed that some packages follow their typical release cycles when releasing security fixes, resulting in large code changes in the release. Even when a package issues a new release immediately after a security fix, the release may still come bundled with all the code changes made since the last release. Overall, our analysis suggests that open source packages do not follow the recommendation of issuing separate releases for security fixes. 

\textbf{Package maintainers may not (always) strictly follow SemVer formatting~\cite{semver} when making breaking changes:} SemVer versioning is a guideline for package maintainers to indicate the code change type in a certain release. However, there is no technical method to measure if the maintainers are accurately following the guideline. Prior work has found that the introduction of breaking changes in non-major releases (in violation of SemVer policy) is prevalent among Maven packages~\cite{raemaekers2014semantic}. In our analysis of release notes, we also find non-major releases to mention breaking changes. Specifically, out of 32 security releases that mention breaking change in the release notes, 6 were a patch release and 12 were a minor release. Including breaking changes in non-major stable releases validates developers' fear of regression when updating dependencies.

\subsection{CVEs vs non-CVEs}

Imtiaz et al.~\cite{imtiaz2021comparative}, when comparing popular SCA tools, found six of the nine studied tools to have reported unique non-CVEs that were not reported by the other tools in the study. However, while the CVE data is widely accepted within the security community, and well studied in the literature~\cite{frei2006large,shahzad2012large, nakajima2019pilot, shahzad2019large, ruohonen2020mixed, li2017large}, vulnerabilities without a CVE identifier are still under-studied, although they are not uncommon in the open source vulnerability databases~\cite{imtiaz2021comparative}. Approximately one-fourth of the advisories in our dataset are non-CVEs which provides us with a unique opportunity to compare CVEs and non-CVEs relative to security release practices.

In the measurement for RQ1, RQ2, and RQ3, we do not find any difference between CVEs and non-CVEs besides our finding that the CVEs are more likely to contain an explicit security mention in the release notes than the non-CVEs. Further, we find that the non-CVEs have a higher delay in being published on advisory databases than the CVEs. As noted before, security researchers and SCA toolmakers often scan open source packages to discover (previously) unreported security fixes, which may explain the higher advisory publication delay.

However, we do not find any evidence in our data that CVEs and non-CVEs get treated differently by the package maintainers, indicating that both the CVEs and non-CVEs are considered to carry similar security risks. Therefore, a standardized process to keep track of security releases in the open source packages should be practiced. Further, vulnerability data should also be open source (instead of commercial databases) and be easily queryable by the client projects, e.g. through native package managers, to prevent any reliance on commercial tools.

\subsection{Comparison between different package ecosystems }
We study security release practices over seven package ecosystems. Regarding fix-to-release delay, we find that Maven overall has a higher delay than the other ecosystems. In our manual analysis, we found that such a delay may occur because the security fix will be announced after the release. The typical release cycle and the testing required before a new release may cause a higher fix-to-release delay in the case of large-scale projects. Further, often multiple packages are maintained together as a single project, and follow a single release schedule which may also impact the fix-to-release delay. We observed the practice of maintaining multiple packages as a single project to be more common in Maven, NuGet and Go. On the other end of the spectrum, the npm ecosystem is known for \textit{micropackages} where packages are small and dedicated to performing single tasks~\cite{abdalkareem2017developers, zimmermann2019small}. While the pros and cons of micropackages can be debated~\cite{kula2017impact}, we observe that the npm packages, at a median,  have security releases within 24 hours of the corresponding fix. 

Similarly, we find higher code change size in security releases in Composer, Maven, NuGet, and pip packages compared to the other three ecosystems. In our manual analysis, we find that the code change size may be determined by the release cycle and the developmental activity of a given package. Therefore, the difference in code change size may be coming from the project size and the release cycles. We observed that Go, npm, and RubyGems packages are often small, and consequently, have small security releases as well.

Regarding documentation, we find that Python packages document security fixes in the release notes 82.8\% of the time, while Maven packages only document security fixes 40.8\% of the time. Further, Go, NuGet and pip packages are the  most likely to have a release note coming with the security releases. When we could actually find a release note, we see that the notes are highly likely to mention the associated security fix (84.6\% of the time). Therefore, the difference may be coming from how the packages actually distribute information on updates, or if they do at all.

Regarding the delay in advisory publication, we discussed the long delay in RubyGems in Section \ref{sec:advpubdelay}. We see NuGet packages to have the least median delay (2 days). Most NuGet packages in our dataset are from the Microsoft organization, which may cause a uniformity in practices. When a vulnerability is discovered by an external source, e.g. bug bounty hunters, an advisory may be published quickly. However, when security fixes are made as part of the regular development, the fix may not get reported in the vulnerability databases even if the vulnerability affects a public release. Future work may investigate if/how packages fix security issues as part of the regular development, and if such fixes are reported to the vulnerability databases at the time of the subsequent releases.

\subsection{Recommendations}
\label{sec:recommendation}
In this section, we provide four recommendations for the package maintainers and the ecosystem administrators for better handling of security releases based on the findings of our study.

\subsubsection{Private forks for security fixes} While we find open source packages to be typically fast in bundling a security fix in a new release, 25\% of the releases in our dataset still come at least 20 days after the corresponding security fix. While there may be valid reasons why packages need time before the next release, e.g., testing, contributor coordination, etc., the commit messages and issue discussion in open source make security fixes easy to spot as we have observed in our manual analysis.

Therefore, a solution to mitigate the security risk of fix-to-release delay is to keep the security fixes private until the next release. At the start of the vulnerability discovery and reporting process, when an external party discovers a vulnerability in a project, maintainers recommend responsible disclosure through a private channel~\cite{nakajima2019pilot}.
Afterward, when working on a fix, GitHub offers a feature where open source package maintainers can create a private fork to collaborate on the fix~\footnote{https://docs.github.com/en/code-security/security-advisories/collaborating-in-a-temporary-private-fork-to-resolve-a-security-vulnerability}. However, once the fix is finalized and merged into the main codebase, the fix, and the associated commit message becomes public. 

We recommend that the private fork for a security fix should remain private until the next release of the project. After completing a security fix in a private fork, developers can approve the changes and indicate the commit point where the security fix will be merged. However, the fix commit(s) should become publicly visible only at the time of the next release, preferably automatically. GitLab, a code hosting platform, follows an open source development model but keeps security fixes private until a release in a similar way~\footnote{https://about.gitlab.com/blog/2021/01/04/how-we-prevented-security-fixes-leaking-into-our-public-repositories/}.

Further, private forks can also help automate tagging a release as a security release and publishing an advisory immediately at the time of the release. On GitHub, developers have to declare a security advisory first to enable the private fork option. This way, package maintainers would only have to enable a private fork when working on security fixes, and the subsequent desired actions can be automated: (i) keeping the fix private until release; (ii) tagging the new release as a security release; and (iii) publishing an advisory.

We recognize that there may be technical challenges in keeping the security fix private until the next release and meanwhile add new changes from collaborators who do not have visibility of the fix. However, private forks may also nudge package maintainers to prioritize a new release upon making security fixes, and consequently, make the fix available sooner to the client projects. 
    
    \subsubsection{Standardized practice for announcing security release} When an open source package issues a new security release, the client projects should get notified immediately. However, manually reporting vulnerability data to advisory databases may create a lag in the process, as shown in RQ4. Further, many security releases may remain unreported. We recommend open source packages follow a standard process in tagging a version as a security release, where the process would also help automate subsequent notifications to the client projects.
    
    A version may be tagged as a security release in multiple possible ways, e.g., (i) add metadata indicating security release in the package manifest files where the version number gets updated; (ii) add a security release label in the release note; or (iii) tag a version commit with standard headers such as \textit{[Security Release]}. A standardized process would make package managers identify security releases automatically and notify the client projects during the clients' next developmental builds. Package managers can also automatically update advisory databases. SCA tools, like Dependabot, can then notify projects without recent activity as well.
    
    Further, in RQ2, we have identified eight categories of information regarding a security fix that get documented in the release notes. Using these categories, ecosystem administrators can create a standardized format that package maintainers should include in the release notes of a security release. Vulnerability databases can leverage the information in the release notes to add a new advisory along with the necessary metadata, and a manual reporting of the advisories would not be required. This way, different SCA tools can incorporate a vulnerability data all at the same time. 
    
    However, we have also observed no standard format for release notes, let alone providing vulnerability information.  We recommend package maintainers write release notes and keep them in a common source similar to README.md used to describe the project, CONTRIBUTING.md. used to explain the contribution policies, SECURITY.md used to explain the responsible disclosure policy, etc. We observe that CHANGELOG.md is a commonly followed file naming format across different languages for the release notes. Ecosystem administrators can decide on a standard format for the release notes and provide badges for packages that provide release notes for each version. Prior work has found that such badges can nudge package maintainers to follow good developmental practices and help client projects to select dependencies~\cite{trockman2018adding}.
    
    \subsubsection{Security patch for older versions} We have discussed the code change size in security releases and the possible regression risk these releases may carry. However, the analysis may be relevant only for the client projects who are using the latest version of a package in a specific release branch. Projects with outdated dependencies -- that is -- using a version of the package older than the latest available may face greater migration effort in upgrading to a security release~\cite{kula2018developers}. Therefore, popular packages may release a security patch, at least for critical vulnerabilities, that can be applied to the past versions of the package. 
    
    For example, if a package has the latest version at 3.2.14, and the security release comes at 3.2.15, the package can also release a security patch that can be applied to any version starting from 3.0.0 (e.g., from 3.0.0 to 3.0.0.1 indicating a security patch adoption). This way, client projects can immediately adopt a security fix in case of regression risks. Snyk, an SCA tool, provides such security patches for selected vulnerabilities for Node.js (JavaScript) projects~\cite{snykpatch}. Future research can look at how to automate such a process -- that is -- (i) identifying the fix commits of a vulnerability; (ii) converting the fix commits into a stand-alone patch; (ii) identify all the older releases where the patch can be automatically applied without any regression; and (iii) apply the patch in the client project and keep a record of the fix. 
    
    \subsubsection{Security metrics for dependency selection} Ecosystem administrators can provide automatically measured security metrics with packages to aid security-critical projects in dependency selection. Prior work has found SCA tools to provide a security rating for packages over simple metrics such as vulnerabilities discovered in the past. The Open Source Security Foundation (OSSF) ~\cite{ossf} provides security health metrics for selected packages, such as if the package uses static analysis testing, code review, signed releases, etc.
    
    How packages dealt with security releases in the past can be another metric that signals the reliability of a package in case a vulnerability is found. Specifically, the metrics in RQ1-4 can be utilized to signal past security release practices of a project. For example, how quickly the fix was released, if the release note mentioned the security fix, and if an advisory was immediately published for the past security releases. Similarly, future research can look at (i) if providing security metrics nudge packages to employ better practices; and (ii) if such metrics help the client projects in dependency selection and maintain their own projects' overall security health.

\section{Threats to validity}
\label{sec:threats}


The primary threat to the validity of our findings is how accurate is our data set, and whether the data generalize beyond the seven ecosystems we study. We provide data and methodology accuracy checks throughout the paper for indication of the reliability of our analysis.
Further, a database may be biased towards certain types of vulnerabilities and software packages depending on the vulnerability reporting and data collection process~\cite{li2017large, christey2013buying}. We address this threat by leveraging the Snyk database that is specifically curated for vulnerabilities in the open source packages and includes non-CVEs and CVEs from NVD. However, we worked only with the publicly available data from the Snyk database, which may introduce unknown biases. Similarly, the analysis for RQ1 incorporates vulnerabilities only where the fix commit is known and may be biased. We report the count of distinct packages and CWE types
while reporting our findings in this study
to indicate the generalizability of our results. 

For the analysis of RQ1, advisory references may point to fix commits applied to different release branches. However, we considered the latest of these commit dates as the fix commit date for all the branches, which may not be accurate in all cases. We believe the threat is minimal, as we find security releases in all the branches are likely to come on the same day, 51.8\% of the time. For RQ2, we explain how we located the release notes. However, we may have missed other announcement platforms, such as Twitter. To ensure the reliability of the qualitative analysis,
two of the authors independently performed the analysis, which took 50 hours for each person.

Further, the additional analyses presented in Section ~\ref{sec:discussion} also have limitations. In measuring advisory publication delay, we only consider two databases (Snyk and NVD) while there exist other popular databases as well, such as GitHub  and npm security advisory databases. Nonetheless, our analysis shows the risk of delayed notification when using Snyk as an SCA tool.

\section{Conclusion }
\label{sec:conclusions}

In this paper, we study fix-to-release delay, documentation, code change characteristics, and advisory publication delay for past security releases across seven open source package ecosystems, namely Composer, Go, Maven, npm, NuGet, pip, and RubyGems. We find that the open source packages are typically fast in releasing security fixes, as the median release comes within 4 days of the corresponding security fix. However, 25\% of the releases still have a delay of at least 20 days. While an advisory may be published only after the security release, the commit messages and issue discussion within the code repository makes it possible for potential attackers to discover the vulnerability and gain a window of opportunity to plan attacks on the client projects.

Further, even after the security release, the client projects may remain unaware of the security fix, as we find a median delay of 25 days before the publication of an advisory on Snyk and NVD. We analyzed the release notes of the security release to investigate if and how the release note documents the security fixes. We found that 38.5\% of the security releases either do not have a release note, or the release note does not mention the security fix. Conversely, we find that even when the release note explicitly mentions the security fixes, 26.7\% of the security releases still take at least 30 days before a corresponding advisory is published. Such a time lag can result in delayed notifications to the client projects through an SCA tool, and can further increase the window of opportunity for the attackers. To address these security threats, we recommend package maintainers work in private forks for security fixes and follow a standardized practice when announcing security releases so that the client projects are notified automatically and immediately. 

Lastly, we find that security releases contain 134 lines of code changes and modify 6 source code files at a median. We observe that the security releases may also come with unrelated functional changes and even breaking changes, as packages typically do not follow the recommendation of dedicating a separate release only for the security fixes. Further, the potential migration effort may be higher for the client projects that are using an older version to the latest available of a dependency package. In this regard, we recommend packages release security patches, at least for critical vulnerabilities, that can be applied to older versions of the package.

To conclude, the reliance on open source packages in modern software development makes the client projects' security health to be directly correlated with the security health of their dependency packages~\cite{cox2019surviving}. Therefore, projects should select dependencies keeping a package's security practices in mind~\cite{pashchenko2020qualitative}. Ecosystem administrators may provide metrics that signal the security health of a package to aid the client projects in this regard. One such signal can be how security releases in the past were handled by the packages. Such a signal will not only help projects in dependency selection, it may also nudge the package maintainers in following better developmental practices. Moreover, packages should employ a standardized format when announcing security releases so that the client projects are notified automatically when a release is available and have less reliance on any third-party vulnerability databases and commercial security tools.

\section{Acknowledgment}
We thank the Realsearch group for providing valuable feedback. Our research was funded by the NSA Science of Security Lablet.

\bibliographystyle{IEEEtran}
\bibliography{bibliography}

\begin{thebibliography}{10}
\providecommand{\url}[1]{#1}
\csname url@samestyle\endcsname
\providecommand{\newblock}{\relax}
\providecommand{\bibinfo}[2]{#2}
\providecommand{\BIBentrySTDinterwordspacing}{\spaceskip=0pt\relax}
\providecommand{\BIBentryALTinterwordstretchfactor}{4}
\providecommand{\BIBentryALTinterwordspacing}{\spaceskip=\fontdimen2\font plus
\BIBentryALTinterwordstretchfactor\fontdimen3\font minus
  \fontdimen4\font\relax}
\providecommand{\BIBforeignlanguage}[2]{{%
\expandafter\ifx\csname l@#1\endcsname\relax
\typeout{** WARNING: IEEEtran.bst: No hyphenation pattern has been}%
\typeout{** loaded for the language `#1'. Using the pattern for}%
\typeout{** the default language instead.}%
\else
\language=\csname l@#1\endcsname
\fi
#2}}
\providecommand{\BIBdecl}{\relax}
\BIBdecl

\bibitem{sonatype2021}
``2021 state of the software supply chain,''
  \url{https://www.sonatype.com/resources/state-of-the-software-supply-chain-2021},
  2021.

\bibitem{blackduck2021}
Synopsys, ``2021 open source security and risk analysis report,''
  \url{https://www.synopsys.com/software-integrity/resources/analyst-reports/open-source-security-risk-analysis.html},
  2021.

\bibitem{cox2019surviving}
R.~Cox, ``Surviving software dependencies,'' \emph{Communications of the ACM},
  vol.~62, no.~9, pp. 36--43, 2019.

\bibitem{zimmermann2019small}
M.~Zimmermann, C.-A. Staicu, C.~Tenny, and M.~Pradel, ``Small world with high
  risks: A study of security threats in the npm ecosystem,'' in \emph{28th
  $\{$USENIX$\}$ Security Symposium ($\{$USENIX$\}$ Security 19)}, 2019, pp.
  995--1010.

\bibitem{ossvuln}
``Open source software vulnerabilities increased by 130\% in 2019,''
  \url{https://www.infosecurity-magazine.com/news/open-source-vulnerabilities},
  2020.

\bibitem{li2017large}
F.~Li and V.~Paxson, ``A large-scale empirical study of security patches,'' in
  \emph{Proceedings of the 2017 ACM SIGSAC Conference on Computer and
  Communications Security}, 2017, pp. 2201--2215.

\bibitem{chinthanet2021lags}
B.~Chinthanet, R.~G. Kula, S.~McIntosh, T.~Ishio, A.~Ihara, and K.~Matsumoto,
  ``Lags in the release, adoption, and propagation of npm vulnerability
  fixes,'' \emph{Empirical Software Engineering}, vol.~26, no.~3, pp. 1--28,
  2021.

\bibitem{pashchenko2020qualitative}
I.~Pashchenko, D.-L. Vu, and F.~Massacci, ``A qualitative study of dependency
  management and its security implications,'' in \emph{Proceedings of the 2020
  ACM SIGSAC Conference on Computer and Communications Security}, 2020, pp.
  1513--1531.

\bibitem{kula2018developers}
R.~G. Kula, D.~M. German, A.~Ouni, T.~Ishio, and K.~Inoue, ``Do developers
  update their library dependencies?'' \emph{Empirical Software Engineering},
  vol.~23, no.~1, pp. 384--417, 2018.

\bibitem{martiusdoes}
F.~Martius and C.~Tiefenau, ``What does this update do to my systems?--an
  analysis of the importance of update-related information to system
  administrators.''

\bibitem{ruohonen2020mixed}
J.~Ruohonen, S.~Hyrynsalmi, and V.~Lepp{\"a}nen, ``A mixed methods probe into
  the direct disclosure of software vulnerabilities,'' \emph{Computers in Human
  Behavior}, vol. 103, pp. 161--173, 2020.

\bibitem{ramsauer2020sound}
R.~Ramsauer, L.~Bulwahn, D.~Lohmann, and W.~Mauerer, ``The sound of silence:
  Mining security vulnerabilities from secret integration channels in
  open-source projects,'' in \emph{Proceedings of the 2020 ACM SIGSAC
  Conference on Cloud Computing Security Workshop}, 2020, pp. 147--157.

\bibitem{bi2020empirical}
T.~Bi, X.~Xia, D.~Lo, J.~Grundy, and T.~Zimmermann, ``An empirical study of
  release note production and usage in practice,'' \emph{IEEE Transactions on
  Software Engineering}, 2020.

\bibitem{keepachangelog}
``keep a changelog,'' \url{https://keepachangelog.com/en/1.0.0/}, 2021.

\bibitem{changelog}
``Changelog,'' \url{https://en.wikipedia.org/wiki/Changelog}, 2021.

\bibitem{semver}
``Semantic versioning 2.0.0,'' \url{https://semver.org/}, 2021.

\bibitem{dependabot}
``Dependabot,'' \url{https://dependabot.com/}.

\bibitem{alfadel2021empirical}
M.~Alfadel, D.~E. Costa, and E.~Shihab, ``Empirical analysis of security
  vulnerabilities in python packages,'' in \emph{2021 IEEE International
  Conference on Software Analysis, Evolution and Reengineering (SANER)}.\hskip
  1em plus 0.5em minus 0.4em\relax IEEE, 2021, pp. 446--457.

\bibitem{imtiaz2021comparative}
N.~Imtiaz, S.~Thorn, and L.~Williams, ``A comparative study of vulnerability
  reporting by software composition analysis tools,'' in \emph{Proceedings of
  the 15th ACM/IEEE International Symposium on Empirical Software Engineering
  and Measurement (ESEM)}, 2021, pp. 1--11.

\bibitem{nvd}
``National vulnerability database,'' \url{https://nvd.nist.gov/vuln}.

\bibitem{snykdb}
``Snyk vulnerability db,'' \url{https://snyk.io/vuln}, 2021.

\bibitem{securityadvisory}
S.~Maddox, ``Writing a security advisory,''
  \url{https://ffeathers.wordpress.com/2010/08/08/writing-a-security-advisory/}.

\bibitem{lauinger2018thou}
T.~Lauinger, A.~Chaabane, S.~Arshad, W.~Robertson, C.~Wilson, and E.~Kirda,
  ``Thou shalt not depend on me: Analysing the use of outdated javascript
  libraries on the web,'' \emph{arXiv preprint arXiv:1811.00918}, 2018.

\bibitem{wang2020empirical}
Y.~Wang, B.~Chen, K.~Huang, B.~Shi, C.~Xu, X.~Peng, Y.~Wu, and Y.~Liu, ``An
  empirical study of usages, updates and risks of third-party libraries in java
  projects,'' in \emph{2020 IEEE International Conference on Software
  Maintenance and Evolution (ICSME)}.\hskip 1em plus 0.5em minus 0.4em\relax
  IEEE, 2020, pp. 35--45.

\bibitem{decan2017empirical}
A.~Decan, T.~Mens, and M.~Claes, ``An empirical comparison of dependency issues
  in oss packaging ecosystems,'' in \emph{2017 IEEE 24th International
  Conference on Software Analysis, Evolution and Reengineering (SANER)}.\hskip
  1em plus 0.5em minus 0.4em\relax IEEE, 2017, pp. 2--12.

\bibitem{salza2019third}
P.~Salza, F.~Palomba, D.~Di~Nucci, A.~De~Lucia, and F.~Ferrucci, ``Third-party
  libraries in mobile apps: when, how, and why developers update
  them-appendix,'' 2019.

\bibitem{zerouali2019formal}
A.~Zerouali, T.~Mens, J.~Gonzalez-Barahona, A.~Decan, E.~Constantinou, and
  G.~Robles, ``A formal framework for measuring technical lag in component
  repositories—and its application to npm,'' \emph{Journal of Software:
  Evolution and Process}, vol.~31, no.~8, p. e2157, 2019.

\bibitem{gonzalez2020characterizing}
J.~M. Gonzalez-Barahona, ``Characterizing outdateness with technical lag: an
  exploratory study,'' in \emph{Proceedings of the IEEE/ACM 42nd International
  Conference on Software Engineering Workshops}, 2020, pp. 735--741.

\bibitem{stringer2020technical}
J.~Stringer, A.~Tahir, K.~Blincoe, and J.~Dietrich, ``Technical lag of
  dependencies in major package managers,'' in \emph{2020 27th Asia-Pacific
  Software Engineering Conference (APSEC)}.\hskip 1em plus 0.5em minus
  0.4em\relax IEEE, 2020, pp. 228--237.

\bibitem{decan2018evolution}
A.~Decan, T.~Mens, and E.~Constantinou, ``On the evolution of technical lag in
  the npm package dependency network,'' in \emph{2018 IEEE International
  Conference on Software Maintenance and Evolution (ICSME)}.\hskip 1em plus
  0.5em minus 0.4em\relax IEEE, 2018, pp. 404--414.

\bibitem{decan2018impact}
------, ``On the impact of security vulnerabilities in the npm package
  dependency network,'' in \emph{Proceedings of the 15th International
  Conference on Mining Software Repositories}, 2018, pp. 181--191.

\bibitem{prana2021out}
G.~A.~A. Prana, A.~Sharma, L.~K. Shar, D.~Foo, A.~E. Santosa, A.~Sharma, and
  D.~Lo, ``Out of sight, out of mind? how vulnerable dependencies affect
  open-source projects,'' \emph{Empirical Software Engineering}, vol.~26,
  no.~4, pp. 1--34, 2021.

\bibitem{hejderup2015dependencies}
J.~Hejderup, ``In dependencies we trust: How vulnerable are dependencies in
  software modules?'' 2015.

\bibitem{foo2019dynamics}
D.~Foo, J.~Yeo, H.~Xiao, and A.~Sharma, ``The dynamics of software composition
  analysis,'' \emph{arXiv preprint arXiv:1909.00973}, 2019.

\bibitem{frei2006large}
S.~Frei, M.~May, U.~Fiedler, and B.~Plattner, ``Large-scale vulnerability
  analysis,'' in \emph{Proceedings of the 2006 SIGCOMM workshop on Large-scale
  attack defense}, 2006, pp. 131--138.

\bibitem{shahzad2012large}
M.~Shahzad, M.~Z. Shafiq, and A.~X. Liu, ``A large scale exploratory analysis
  of software vulnerability life cycles,'' in \emph{2012 34th International
  Conference on Software Engineering (ICSE)}.\hskip 1em plus 0.5em minus
  0.4em\relax IEEE, 2012, pp. 771--781.

\bibitem{nakajima2019pilot}
A.~Nakajima, T.~Watanabe, E.~Shioji, M.~Akiyama, and M.~Woo, ``A pilot study on
  consumer iot device vulnerability disclosure and patch release in japan and
  the united states,'' in \emph{Proceedings of the 2019 ACM Asia Conference on
  Computer and Communications Security}, 2019, pp. 485--492.

\bibitem{shahzad2019large}
M.~Shahzad, M.~Z. Shafiq, and A.~X. Liu, ``Large scale characterization of
  software vulnerability life cycles,'' \emph{IEEE Transactions on Dependable
  and Secure Computing}, vol.~17, no.~4, pp. 730--744, 2019.

\bibitem{zerouali2021impact}
A.~Zerouali, T.~Mens, A.~Decan, and C.~De~Roover, ``On the impact of security
  vulnerabilities in the npm and rubygems dependency networks,'' \emph{arXiv
  preprint arXiv:2106.06747}, 2021.

\bibitem{alfadel2021use}
M.~Alfadel, D.~E. Costa, E.~Shihab, and M.~Mkhallalati, ``On the use of
  dependabot security pull requests,'' in \emph{2021 IEEE/ACM 18th
  International Conference on Mining Software Repositories (MSR)}.\hskip 1em
  plus 0.5em minus 0.4em\relax IEEE, 2021, pp. 254--265.

\bibitem{imtiaz2019developers}
N.~Imtiaz, B.~Murphy, and L.~Williams, ``How do developers act on static
  analysis alerts? an empirical study of coverity usage,'' in \emph{2019 IEEE
  30th International Symposium on Software Reliability Engineering
  (ISSRE)}.\hskip 1em plus 0.5em minus 0.4em\relax IEEE, 2019, pp. 323--333.

\bibitem{moreno2016arena}
L.~Moreno, G.~Bavota, M.~Di~Penta, R.~Oliveto, A.~Marcus, and G.~Canfora,
  ``Arena: an approach for the automated generation of release notes,''
  \emph{IEEE Transactions on Software Engineering}, vol.~43, no.~2, pp.
  106--127, 2016.

\bibitem{abebe2016empirical}
S.~L. Abebe, N.~Ali, and A.~E. Hassan, ``An empirical study of software release
  notes,'' \emph{Empirical Software Engineering}, vol.~21, no.~3, pp.
  1107--1142, 2016.

\bibitem{snykinfo}
B.~Catabi-Kalman, ``Why do organizations trust snyk to win the open source
  security battle?''
  \url{https://snyk.io/blog/why-snyk-wins-open-source-security-battle/}, 2021.

\bibitem{cwe}
``Common weakness enumeration,'' \url{https://cwe.mitre.org/}.

\bibitem{cve}
``Common vulnerabil- ities and exposures,'' \url{https://cve.mitre.org/cve/},
  2021.

\bibitem{sonatypeapi}
``Sonatype maven central repository search,'' \url{https://search.maven.org/}.

\bibitem{git}
``Git --local-branching-on-the-cheap,'' \url{https://git-scm.com/}.

\bibitem{svn}
``Apache subversion,'' \url{https://subversion.apache.org/}.

\bibitem{githubviewablecommit}
``Commit exists on github but not in my local clone,''
  \url{https://docs.github.com/en/github/committing-changes-to-your-project/commit-exists-on-github-but-not-in-my-local-clone#the-branch-that-contained-the-commit-was-deleted}.

\bibitem{gomodule}
``Go modules,''
  \url{https://github.com/golang/go/wiki/Modules#publishing-a-release}.

\bibitem{packaging}
``Packaging,'' \url{https://pypi.org/project/packaging/}.

\bibitem{mavensemver}
``mvn-compare,'' \url{https://pypi.org/project/mvn-compare/}.

\bibitem{mavenversioncompare}
``Class comparableversion,''
  \url{https://maven.apache.org/ref/3.3.3/maven-artifact/apidocs/org/apache/maven/artifact/versioning/ComparableVersion.html}.

\bibitem{nuget}
``An introduction to nuget,''
  \url{https://docs.microsoft.com/en-us/nuget/what-is-nuget}.

\bibitem{mann1947test}
H.~B. Mann and D.~R. Whitney, ``On a test of whether one of two random
  variables is stochastically larger than the other,'' \emph{The annals of
  mathematical statistics}, pp. 50--60, 1947.

\bibitem{weisstein2004bonferroni}
E.~W. Weisstein, ``Bonferroni correction,'' \emph{https://mathworld. wolfram.
  com/}, 2004.

\bibitem{embargo}
``The hidden costs of embargoes,''
  \url{https://access.redhat.com/blogs/766093/posts/1976653}, 2015.

\bibitem{bogart2016break}
C.~Bogart, C.~K{\"a}stner, J.~Herbsleb, and F.~Thung, ``How to break an api:
  cost negotiation and community values in three software ecosystems,'' in
  \emph{Proceedings of the 2016 24th ACM SIGSOFT International Symposium on
  Foundations of Software Engineering}, 2016, pp. 109--120.

\bibitem{khandkar2009open}
S.~H. Khandkar, ``Open coding,'' \emph{University of Calgary}, vol.~23, p.
  2009, 2009.

\bibitem{hancock2001introduction}
B.~Hancock, E.~Ockleford, and K.~Windridge, \emph{An introduction to
  qualitative research}.\hskip 1em plus 0.5em minus 0.4em\relax Trent focus
  group, 2001.

\bibitem{wicks2017coding}
D.~Wicks, ``The coding manual for qualitative researchers,'' \emph{Qualitative
  research in organizations and management: an international journal}, 2017.

\bibitem{viera2005understanding}
A.~J. Viera, J.~M. Garrett \emph{et~al.}, ``Understanding interobserver
  agreement: the kappa statistic,'' \emph{Fam med}, vol.~37, no.~5, pp.
  360--363, 2005.

\bibitem{plackett1983karl}
R.~L. Plackett, ``Karl pearson and the chi-squared test,'' \emph{International
  Statistical Review/Revue Internationale de Statistique}, pp. 59--72, 1983.

\bibitem{zhou2017automated}
Y.~Zhou and A.~Sharma, ``Automated identification of security issues from
  commit messages and bug reports,'' in \emph{Proceedings of the 2017 11th
  joint meeting on foundations of software engineering}, 2017, pp. 914--919.

\bibitem{jeremy_katz_2020_3626071}
\BIBentryALTinterwordspacing
J.~Katz, ``{Libraries.io Open Source Repository and Dependency Metadata},''
  Jan. 2020. [Online]. Available: \url{https://doi.org/10.5281/zenodo.3626071}
\BIBentrySTDinterwordspacing

\bibitem{sedgwick2014spearman}
P.~Sedgwick, ``Spearman’s rank correlation coefficient,'' \emph{Bmj}, vol.
  349, 2014.

\bibitem{akoglu2018user}
H.~Akoglu, ``User's guide to correlation coefficients,'' \emph{Turkish journal
  of emergency medicine}, vol.~18, no.~3, pp. 91--93, 2018.

\bibitem{equifax}
J.~Fruhlinger, ``Equifax data breach faq: What happened, who was affected, what
  was the impact?''
  \url{https://www.csoonline.com/article/3444488/equifax-data-breach-faq-what-happened-who-\\was-affected-what-was-the-impact.html},
  2020.

\bibitem{raemaekers2014semantic}
S.~Raemaekers, A.~Van~Deursen, and J.~Visser, ``Semantic versioning versus
  breaking changes: A study of the maven repository,'' in \emph{2014 IEEE 14th
  International Working Conference on Source Code Analysis and
  Manipulation}.\hskip 1em plus 0.5em minus 0.4em\relax IEEE, 2014, pp.
  215--224.

\bibitem{abdalkareem2017developers}
R.~Abdalkareem, O.~Nourry, S.~Wehaibi, S.~Mujahid, and E.~Shihab, ``Why do
  developers use trivial packages? an empirical case study on npm,'' in
  \emph{Proceedings of the 2017 11th joint meeting on foundations of software
  engineering}, 2017, pp. 385--395.

\bibitem{kula2017impact}
R.~G. Kula, A.~Ouni, D.~M. German, and K.~Inoue, ``On the impact of
  micro-packages: An empirical study of the npm javascript ecosystem,''
  \emph{arXiv preprint arXiv:1709.04638}, 2017.

\bibitem{trockman2018adding}
A.~Trockman, S.~Zhou, C.~K{\"a}stner, and B.~Vasilescu, ``Adding sparkle to
  social coding: an empirical study of repository badges in the npm
  ecosystem,'' in \emph{Proceedings of the 40th International Conference on
  Software Engineering}, 2018, pp. 511--522.

\bibitem{snykpatch}
``Snyk patches to fix vulnerabilities,''
  \url{https://docs.snyk.io/fixing-and-prioritizing-issues/starting-to-fix-vulnerabilities/snyk-patches-to-fix-vulnerabilities}.

\bibitem{ossf}
``Security scorecards for open source projects,''
  \url{https://openssf.org/blog/2020/11/06/security-scorecards-for-open-source-projects/}.

\bibitem{christey2013buying}
S.~Christey and B.~Martin, ``Buying into the bias: Why vulnerability statistics
  suck,'' \emph{BlackHat, Las Vegas, USA, Tech. Rep}, vol.~1, 2013.

\end{thebibliography}
\end{document}